\documentclass[a4paper,11pt]{article}
\usepackage{jcappub}
\usepackage{caption} % to edit figure numbering
\usepackage{xcolor}
\usepackage{ulem}
\begin{document}

\title{Directional direct detection of light dark matter up-scattered by cosmic rays from direction of the Galactic center }

\author[a,1]{Keiko~I.~Nagao}
\author[b,2]{Satoshi~Higashino}
\author[c,3]{Tatsuhiro~Naka}
\author[c,4]{Kentaro~Miuchi}

\affiliation[a]{%
 Department of Physics, Okayama University of Science, Okayama, 700-0005, Japan.
}%
\affiliation[b]{%
 Department of Physics, Kobe University, Hyogo 657-8501, Japan.
}%
\affiliation[c]{%
Department of Physics, Toho University, Chiba 274-8510, Japan.%,
%Kobayashi-Maskawa Institute, Nagoya University, Aichi 464-8602, Japan.
}%

\emailAdd{nagao@ous.ac.jp}
\emailAdd{higashino@people.kobe-u.ac.jp}
\emailAdd{tatsuhiro.naka@sci.toho-u.ac.jp}
\emailAdd{miuchi@phys.sci.kobe-u.ac.jp}

             %  but any dat\affiliation[c]{%

%e may be explicitly specified

\abstract{
Dark matter with MeV scale mass is difficult to detect with standard direct search detectors. 
However, they can be searched for by considering the up-scattering of kinetic energies by cosmic rays. Because the dark matter density is higher in the central region of the Galaxy, the up-scattered dark matter will arrive at Earth from the direction of the Galactic center. Once the dark matter is detected, we can expect to recognize this feature by directional direct detection experiments. In this study, we simulate the nuclear recoils of the up-scattered dark matter and 
%by cosmic rays
quantitatively reveal that a large amount of this type of dark matter is arriving from the direction of the Galactic center. %Also, we define a parameter that indicates that dark matter is coming from the direction of the Galactic center as expected. By numerical simulation and analysis using the parameter, we show that it can be verified by directional detectors with future upgrades.
Also, we have shown that the characteristic signatures %of CR-DMs 
 of the up-scattered dark matter can be verified with more than 5$\sigma$ confidence levels for the assumed target atoms and future upgrades to directional detectors.
}

%\pacs{\textcolor{blue}{Valid PACS appear here}}% PACS, the Physics and Astronomy

\maketitle
\flushbottom
%%%%%%%%%%%%%%%%%%%%
\section{Introduction}\label{sec:introduction}
%%%%%%%%%%%%%%%%%%%%
The identity of dark matter (DM) is a highly significant question in cosmology and particle physics. In the past a few decades, weakly interacting massive particles (WIMPs) have been expected to be the identity of DM. Many attempts have been made to detect WIMPs directly and indirectly, and to produce them in collider experiments. However, despite advances in the sensitivity of  direct detection experiments, no obvious detection of DM has been achieved so far. Particularly, direct detection experiments have placed very stringent limits on the magnitude of the interaction cross section for WIMP particles 
of GeV-scale mass, i.e., $\sigma_\mathrm{SI} < 6.5\times 10^{-48}$~cm$^{2}$ and $\sigma_\mathrm{SD} < 3.1\times 10^{-41}$~cm$^{2}$, where $\sigma_\mathrm{SI (SD)}$ represents spin-independent (SI) and spin-dependent (SD) cross sections of the DM and proton scattering, respectively~\cite{XENON:2018voc, LZ:2022ufs}. On the other hand, for DM having mass lighter than the GeV-scale, it is not easy to obtain high sensitivity through DM-nucleon scatterings. It has recently been proposed that even in such light DM cases, unavoidable scattering between DM and a cosmic ray produces substantial and very energetic DM that can be detected in neutrino detectors and direct detection experiments of DM~\cite{Yin:2018yjn,  Bringmann:2018cvk, Ema:2018bih, Cappiello:2019qsw, Dent:2019krz, Guo:2020drq, Ge:2020yuf, Jaeckel:2020oet, Guo:2020oum, Bell:2021xff, Nagao:2021rio, Feng:2021hyz, Xia:2021vbz, PandaX-II:2021kai, CDEX:2022fig, Xia:2022tid, Super-Kamiokande:2022ncz, Elor:2021swj, Bardhan:2022ywd, Alvey:2022pad, Maity:2022exk}. As well as the up-scattering of DM by cosmic rays, its effect on cosmic ray propagation in the Earth~\cite{Cappiello:2019qsw} has also been investigated. The cosmic ray scattered dark matter (CR-DM) flux has a substantial contribution from the direction of the Galactic center \cite{Cappiello:2019qsw, Ge:2020yuf, Xia:2022tid}. 
In some direct searches such as PandaX and CDEX, constraints for CR-DM have already been placed using real data \cite{PandaX-II:2021kai, CDEX:2022fig}.

Directional direct detection of DM is expected to be very well suited to detect the characteristic events of CR-DM coming from the Galactic center. 
Using directional information to understand the DM nature has been investigated in both directional and non-directional direct searches \cite{Alenazi:2007sy, OHare:2014nxd, Kavanagh:2015aqa, Kavanagh:2016xfi, Catena:2017wzu, Nagao:2017yil, PROSPECT:2021awi}.
Also, verifications of CR-DM using directional information through diurnal modulation in the ordinary direct detection experiments \cite{Cappiello:2019qsw, PandaX-II:2021kai}, and directional search in the neutrino detector \cite{Super-Kamiokande:2022ncz} are already studied.
In this study, we simulate the CR-DM signals in directional searches and reveals its potential for validation of CR-DM.
In directional detection, gas detectors are sensitive to the SD interactions between DM particles and nucleons~\cite{Mayet:2016zxu, Vahsen:2020pzb}. The gases used in the experiments are CF$_4$ and SF$_6$, and in most cases, the nucleon target for scattering with DM is fluorine (F). We also simulate a solid detector NEWSdm~\cite{NEWS:2016fyf, NEWSdm:2017efa}, which is a nuclear emulsion detector and is sensitive to SI interactions. Nuclear emulsion detectors contain multiple targets, thus in the numerical calculations we employ hydrogen ($p$), which is sensitive to light DM, and silver (Ag), which is sensitive to heavy DM.

The rest of this paper is organized as follows. Section~\ref{sec:NumericalSimulation} introduces the methods of the CR-DM simulations and three assumed DM profiles. Section~\ref{sec:numericalresult} presents numerical results and discusses the detectability. Finally, we conclude the paper in Section~\ref{sec:Conclusion}. 

%%%%%%%%%%%%%%%%%%%%
\section{CR-DM and the density profile}
\label{sec:NumericalSimulation}
%%%%%%%%%%%%%%%%%%%%
To simulate the scattering of CR-DM and the targets in direct detection experiments, first a parametrization of the incoming flux of CR-DM to the Earth is required. In this section, we  follow the methods of reference~\cite{Bringmann:2018cvk} to calculate the incoming flux of CR-DM from all directions in space. The calculation procedure is listed below.
%We describe our assumptions, the method of the numerical simulations, and analysis. 

% I suggest using the Latex enviroment \begin{enumerate} \item \end{enumerate}, because it is easier to see when the list of the calculation procedure is finished.  You can also include hyperlinks to different items in the list.
\begin{enumerate}
%1).
  \item If we assume the DM mass $m_\chi$, the interaction cross section $\sigma_{\chi i}$ between DM and the nucleus $i$, density profile $\rho_\chi (r)$, and cosmic ray flux $d\Phi_{i}/dT_{i}$ in the Galaxy, the incoming CR-DM flux can be obtained by
\begin{equation}
\frac{d\Phi_\chi}{dT_\chi d\Omega}=\int_{l.o.s.}\hspace{-5mm} l^2 dl \frac{1}{4\pi l^2} \sigma_{\chi i} \frac{\rho_\chi (r)}{m_\chi} \int_{T_i^\mathrm{min}}^\infty dT_i 
\frac{d{\Phi_i}}{dT_i},
\end{equation}
where $l$ is the length of the line-of-sight (l.o.s) integral, $r$ is the distance from the Galactic center. $T_{\chi}$ and $T_{i}$ are the kinetic energies of the CR-DM and nucleus, respectively. Minimum kinetic energy $T_i^\mathrm{min}$ is defined as
\begin{equation*}
T_i^\mathrm{min} = \left(\frac{T_\chi}{2}-m_i\right)\left(1\pm\sqrt{1+\frac{2T_\chi}{m_\chi}\frac{(m_i+m_\chi)^2}{(2m_i-T_\chi)^2}}\right)
\end{equation*}
where $+$ ($-$) corresponds to the case for $T_\chi>2m_i$ ($T_\chi<2m_i$).
Additionally, we use Hydrogen, which is the most abundant element in space, to be the cosmic ray particle $i$, i.e.,  we assume $i=p$ for simplicity in this paper.

To define the direction of the incoming CR-DM and the nuclear recoils, we adopt the coordinate system shown in Figure~\ref{fig:coordinatesystem.pdf}. We define the direction of the incoming CR-DM with $\theta_\mathrm{DM}$ and $\phi_\mathrm{DM}$, and the direction of the nuclear recoils as $\theta_\mathrm{N}$ and $\phi_\mathrm{N}$. In this coordinate system, $\theta_\mathrm{DM}=\phi_\mathrm{DM}=0$ corresponds to events coming from the direction of the Galactic center.
The scattering angle $\gamma_{\mathrm{G.C.}}$ is the angle from the line connecting the Galactic center to the observer. %, with $\gamma_{\mathrm{G.C.}} = 0$ corresponding to naive scattering from the direction of the Galactic center. 
There is a relation $\cos{\gamma_\mathrm{G.C.}=\cos{\theta_\mathrm{N}} \cos{\phi_\mathrm{N}}}$ between $\theta_\mathrm{N}, \phi_\mathrm{N}$ and $\gamma_\mathrm{G.C.}$.
To include contributions near the Galactic center in the integration of $l$, we take the limits of integration up to the distance between the solar system and the Galactic center $d$ and take $d=8.122$~kpc \cite{de_Salas_2019}\footnote{Although the integration volume might be larger for a more realistic estimation, we limited the range of the l.o.s. integration to 8.122 kpc as a first step of the study. This can be justified as a conservative estimation by having confirmed the DM flux from the Galactic center increase of 40\% (80\%) if we change the range of the line-of-sight integration to 10 kpc (38 kpc).}. 

For the numerical calculation, we use the cosmic ray flux data from the Galprop codebase~\cite{Vladimirov_2011}. Both the cosmic ray sources and the DM profile would be denser in the vicinity of the Galactic center in a realistic scenario; however, following reference~\cite{Bringmann:2018cvk} we assume the flux of cosmic rays is uniform throughout the Galaxy. The dependence of CR-DM flux on cosmic ray profiles near the Galactic Center will be the scope of our future study.
\begin{figure}[htbp] %  figure placement: here, top, bottom, or page
   \centering
   \includegraphics[keepaspectratio, width=0.8\linewidth]{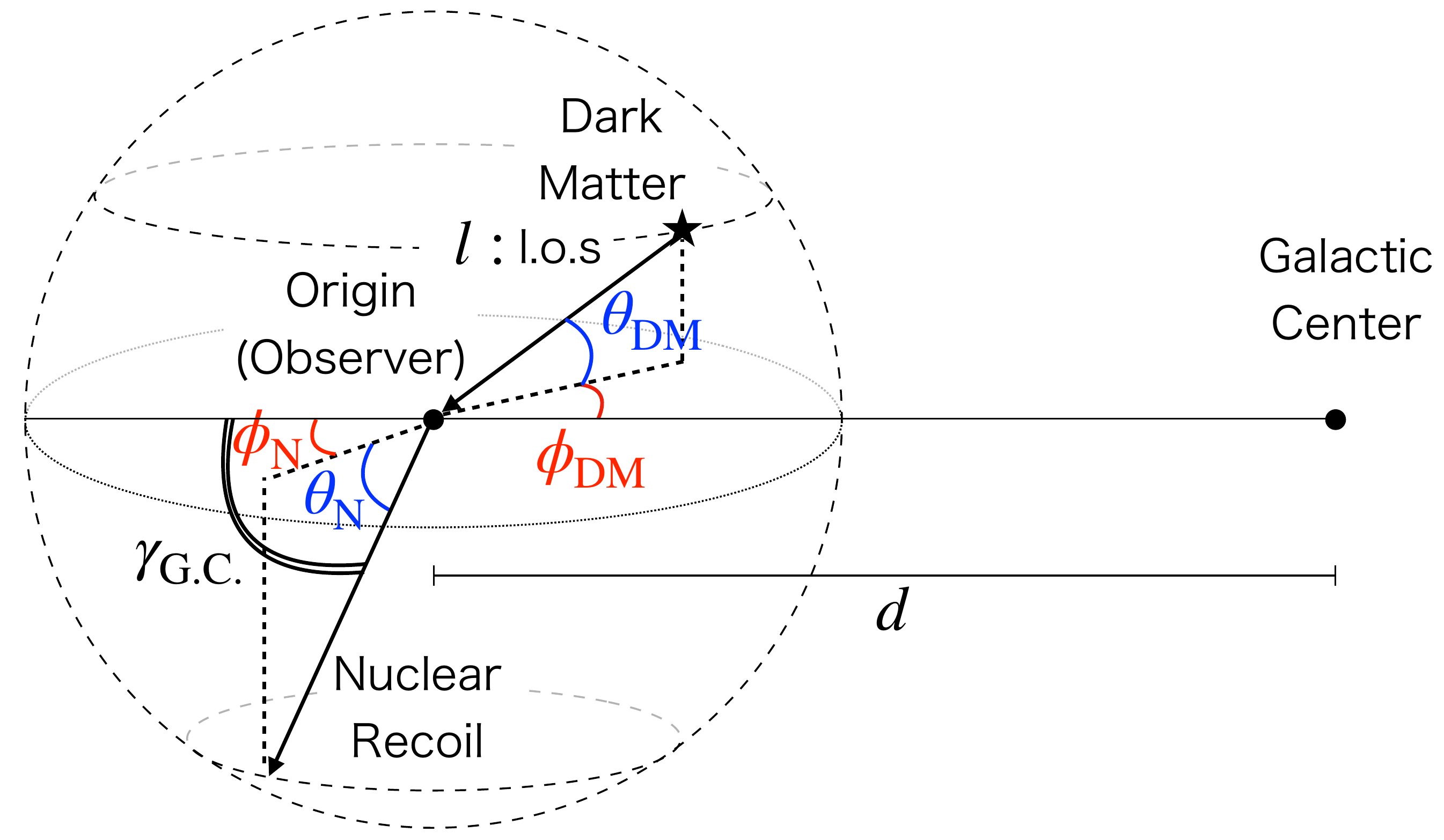} 
   \caption{Coordinate system used in the figures showing the numerical results.}%Coordinate system used in the figures showing the results of the numerical calculations.}
   \label{fig:coordinatesystem.pdf}
\end{figure}
% 2).
  \item\label{en:step2} The obtained CR-DM incoming flux in the Galactic rest frame is transformed to the laboratory frame in the Solar system.
% by ${\bf v}_\textrm{E}={\bf v}_\textrm{G}+{\bf v}_\textrm{Solar}$ where ${\bf v}_\textrm{Solar}=(0, 0, 230)$ km/s. 

% 3).
  \item DM-nucleon scattering caused by the CR-DM in the Solar system is simulated using the CR-DM flux derived in step \ref{en:step2}. Then we obtain the nuclear recoil energy $E_R$ and its scattering angle $\gamma_\mathrm{G.C.}$. 
\end{enumerate}

We assume three typical density profiles for $\rho_\chi (r)$. 
%three typical profiles are assumed.
The first is the frequently used Navarro–Frenk–White (NFW) profile suggested by N-body simulations~\cite{Navarro:1996gj},
\begin{align}
\rho_{\mathrm{NFW}} (r)=\frac{\rho_0}{\frac{r}{r_s}\left(1+\frac{r}{r_s}\right)^2}.
\end{align}
%where %$\rho_s=6.17\times 10^7 M_{\odot}/$kpc$^3$
%$\rho_0=0.403$ GeV$/$cm$^3$, the scale radius %$r_s=7.63$ 
%$r_s=12.53$ kpc, respectively~\cite{Sofue_2012}. 
The second is the Einasto profile which gives a better fit to the N-body simulation data than the NFW profile~\cite{Navarro_2004, Graham_2006}. It is given by 
\begin{align}
\rho_{\mathrm{Ein}}(r)=\rho_0\exp{\left[-\frac{2}{\alpha}\left\{\left(\frac{r}{r_s}\right)^\alpha-1\right\}\right]}.
\end{align}
%where $\alpha=0.17$~\cite{Bernal:2011pz}. 
Because both the NFW and Einasto profiles are cuspy in the Galactic center, we also investigate a cored profile to see the profile dependence clearly. The third profile is the pseudo-isothermal profile (PIT)~\cite{Jimenez:2002vy}
\begin{align}
\rho_{\mathrm{Iso}}(r)=\frac{\rho_0}{1+\left(r/r_s\right)^2}.
\end{align}
Parameters used in the numerical calculation are shown in Table~\ref{Tab:profile}, which are taken from~\cite{de_Salas_2019} for the NFW and the Esnasto profiles, and~\cite{Wegg_2019} for the PIT profile. 
In the table, the local DM density is represented as $\rho_{\mathrm{DM}, \odot}$, and $\rho_0$ is normalized to reproduce that value of $\rho_{\mathrm{DM}, \odot}$.

\begin{table}[thb]
\centering
  \caption{Parameters on DM profiles}
  \begin{tabular}{l||c|c|c|c}  
    & $\rho_{\mathrm{DM}, \odot}$  [GeV/cm$^3$] & $\rho_0$ [GeV/cm$^3$] & $r_s$  [kpc] & $\alpha$ \\ \hline \hline
    NFW & 0.38 & 0.83 & 11.0 & - \\ \hline
    Einasto & 0.38 & 0.30 & 9.2 & 0.18 \\ \hline
    PIT & 0.35 & 3.56 & 2.7 & - \\ 
  \end{tabular}
  \label{Tab:profile}
\end{table}

%%%%%%%%%%%%%%%%%%%%
\section{Numerical Result and discussion}
\label{sec:numericalresult}
%%%%%%%%%%%%%%%%%%%%
In direct detection experiments with directional sensitivity, the energy and angle of nuclear recoils are expected to be detected.
%The three cases of target: hydrogen (p), fluorine (F), and silver (Ag) are assumed.
We will discuss the results obtained from simulations
% each of these
in this section.

%%%%%%%%%%%%%%%%%%%%
\subsection{Angular distribution of CR-DM and nuclear recoil}
\label{subsec:skymap}
%%%%%%%%%%%%%%%%%%%%
Figure~\ref{Fig:skymap_dm} shows the
sky maps for the incoming distributions 
of the CR-DMs of 10~MeV mass expected in the Solar system for the three density profiles assumed.
%CR-DM in the Solar system for assumed density profiles.
The left, middle, and right columns correspond to the NFW, the Einasto, and the PIT profiles, respectively. In the NFW and Einasto profiles, the incoming flux of CR-DM is highly concentrated in the direction of the Galactic center. Whereas in the PIT profile, the incoming flux of CR-DM is found over a wide area centered at the Galactic center. There is almost no difference in the directional distributions of the CR-DM in the mass range of 1~GeV to 1~MeV.

Next, we move to the directional distributions of nuclear recoils by the CR-DMs. In Figures~\ref{Fig:skymap_p}-\ref{Fig:skymap_Ag}, the angular distributions of the recoiled nuclei are shown in whole sky map density plots for the following reference cases:
\begin{itemize}
\item the targets are $p$, F, and Ag,
\item the DM profiles are NFW, Einasto, and PIT, 
\item the DM masses are 100~MeV, 10~MeV, and 1~MeV.
\end{itemize} 
These figures are intended to illustrate an ideal situation without any consideration of detector response (energy threshold, energy resolution, quenching factor and others) and any types of background.
%In the figures, only nuclear recoil events are shown; that is, background events in direct detection experiments are not included for simplicity. 
%To get general, detector-independent overall pictures the current detector energy thresholds are not imposed, thus the figures include low-momentum events whose energy is smaller than the energy threshold of detectors.
The center of the figures corresponds to the arrival from the direction of the Galactic center, thus in all simulations, the CR-DM flux from the Galactic center is significantly higher than the contribution from other directions.
In the whole sky map density plots, there is no noticeable difference between the simulations using the NFW and Einasto profiles. However, there is a distinct difference between the simulations using the PIT profile and the other two profiles. The different appears because, the NFW and Einasto profiles in the Galactic center are cuspy, but it is cored in the PIT profile. This cuspy vs cored property is reflected in the fact that events are less dense in the Galactic center for simulations using the PIT profile than the other simulations using the NFW and Einasto profiles.

%%%%%%%%%%%%%%%%%%%%
\subsection{Energy dependence of the angular distribution}
\label{subsec:energydependence}
%%%%%%%%%%%%%%%%%%%%
We now discuss the distributions of the nuclear recoil energies and scattering angles to see the dependence of the CR-DM mass, which is not visualized in the sky maps in subsection \ref{subsec:skymap}. % You have a label for the subsection, so I suggest using them to help the reader.
The differential angular distributions  
expected with the $p$, F, and Ag targets are shown in 
Figure~\ref{Fig:recoil_cosGCSI_p},~\ref{Fig:recoil_cosGCSD_F}, and~\ref{Fig:recoil_cosGCSI_Ag}, respectively.
The rate of detection is shown in units of event number per second per kilogram.
%figures show the 
NFW, Einasto and PIT profile simulations are shown in the left, center, and right columns of each figure.
Simulations with a CR-DM mass of 100~MeV, 10~MeV, and 1~MeV are shown in the top, center, and bottom rows of each figure.
%in the directional direct detection are shown.
SI interactions with the DM particles are assumed for the $p$ and Ag targets, while SD interactions are assumed for the F target.
See Appendix \ref{append:SIandSDcrossections} for the sensitivity with the F target for SI interactions.
%Also, as both the spin-independent and spin-dependent cross section, to estimate the number of events in the direct detection experiments, 
We assume the cross section of the DM and the proton interaction is $\sigma_{\chi-p}=10^{-32}$~cm$^2$, which %XENON
is currently not experimentally constrained for SI or SD interactions with $m_\chi \lesssim 0.1$~GeV~\cite{Bringmann:2018cvk}. 
%Fluorine is the target used in the directional direct detection through SD interactions, therefore, we only need to divide by a factor $\eta_A$\footnote{See Appendix for details.} below to derive the number of events in the SD interaction case from the SI interaction case in the figure.

A detector energy threshold of $E_\mathrm{thr}=10$~keV is assumed for the $p$ and F targets, and %a lower limit 
threshold of $E_\mathrm{thr}=100$~keV for the Ag target is assumed. % as the lower limit of the detection energy.
The event rates expected %considering only 
in the energy range greater than the detector energy threshold and which elastic scattering occurs are shown in the figures~\cite{Ikeda_2020}.
%These energy spectra should be compared with the energy spectra of the normal WIMPs...
The black histograms show the distribution over the whole energy range of interest, while colored ones show the distributions corresponding to the recoil energy ranges. We have discarded large momentum transfer events before making these angular distributions, i.e., events with energies greater than 10, 3, and 0.6~MeV are discarded for $p$, F, and Ag targets, respectively. 
%to conservatively focus on the elastic scattering events in this study. 
%the number of events in the plots is estimated to lower than the actual number expected because the high-momentum events are conservatively estimated. 
%In this study, pure elastic scatterings are assumed as the scattering between DM and target nuclei. 
%Actually, 
The reason large momentum transfer events are discarded is that %This is because %it is expected to shift to 
inelastic scatterings are expected to dominate once the kinetic energy of the DM is large enough to see the internal structure of the nucleus. 
To discuss inelastic scatterings between the DM and target nuclei, we would need to introduce some assumptions depending on the type of DM and its interactions~\cite{Ellis:1988nb,Engel:1999kv, Arcadi:2019hrw, Guo:2020oum}. In this paper, we prefer to discuss the properties of the CR-DM in general manners as much as possible, thus we focus only on the elastic scatterings.
%, which can be discussed as a  more plausible approach.
We assume that if the de Broglie wavelength of the DM is longer than that of the target nuclei, it can be treated as a perfect elastic scattering.  For the sake of consistency and considering current detector setups, only such events are included here, thus we conservatively focus only on the elastic scatterings.

As can be easily expected, forward scattering events make peaks
%nuclei are most likely to be recoiled to the direction where they are hit by DM arriving from the direction of the Galactic center, which makes a peak 
at $\cos\gamma_\mathrm{G.C.}=1$, and some distributions have peaks in the center at $\cos\gamma_\mathrm{G.C.}=0$. 
%for some cases.
%light DM, as is more noticeable for  proton and fluorine targets.
These central peaks are clearly visible in the distributions for heavy CR-DMs and disappear in ones for light CR-DMs. %are light.
%Considering scattering in a center-of-mass system, the 
The central peaks correspond to the instance where the target and CR-DM do not change the direction of their motions from their initial ones in the center-of-mass frame through scattering.
This kind of scattering, in which the target and CR-DM graze each other in the center-of-mass frame, becomes less visible in the laboratory frame as the CR-DM mass decreases. %In that case, 
For light CR-DM, in the center-of-mass frame, the CR-DM hits a light target and both the CR-DM and the target are bounced away in opposite directions from its initial state. As a result, a prominent peak is produced at $\cos{\gamma_\mathrm{G.C.}} = 1$, which is the direction expected to be naively scattered from the Galactic center.

The energy dependence is similar for the simulations of NFW and Einasto profiles. On the other hand, in the figures there is a difference between the simulations using the PIT profile and the other two profiles, because of the different treatment of structures %of the profiles
in the center of the galaxy.
%The DM mass also causes a difference in the energy dependence of the angular distribution. %Compared to the case of heavier DM, events tend to be more concentrated in the $\cos{\theta_{\mathrm{GC}}}=1$ direction for lighter DM, but the number of detectable events is in general smaller. 
%Now we will take a closer look at the case of NFW and Einasto profiles. 
%For cases where the DM has large mass, i.e., the upper case, there is a peak at $\cos{\gamma_\mathrm{G.C.}} = 0$. This is the case that DM skims the target, corresponding to scattering with little energy transfer. 
Now we will take a closer look at the simulations using the PIT profile.
The central peak near $\cos{\gamma_\mathrm{G.C.}} = 0$ is moderate and less noticeable than in the NFW and Einasto profiles.
This is because the PIT profile has a core at the Galactic center.  It is clear from Figures~\ref{Fig:skymap_p}-\ref{Fig:skymap_Ag} that the CR-DM signal comes from a wider range than that in the simulations using the other two profiles, in which the CR-DM comes from a narrow range near the Galactic center. 

%%For scattering events with large energy transfers, the recoil nuclei have large kinetic energies, so the events are cut off by the elastic scattering conditions described above. Hence, in these cases, the number of events  used in the analysis is significantly reduced.
%%It is seen that the total number of events 
%%and the fraction of %increases and the number of 
%%low-energy recoil events 
%in the total events also 
%%both increase as the CR-DM becomes lighter. In fact, there are a number of high-energy recoil events for heavy CR-DM cases, which are not included in the analysis because they are not elastic scattering.

\subsection{Asymmetry and sensitivity}
%%%%%%%%%%%%%%%
%%%Asymmetry%%%
%%%%%%%%%%%%%%%

%%%%%%%%%%%%%%%%%%%%
%Asymmetry-DM mass plots
%%%%%%%%%%%%%%%%%%%%
\setcounter{figure}{8}
\begin{figure*}[t]
 \begin{tabular}{cc}
    \centering
    \begin{minipage}[b]{0.5\linewidth}
    \centering
    \includegraphics[keepaspectratio, width=\linewidth]{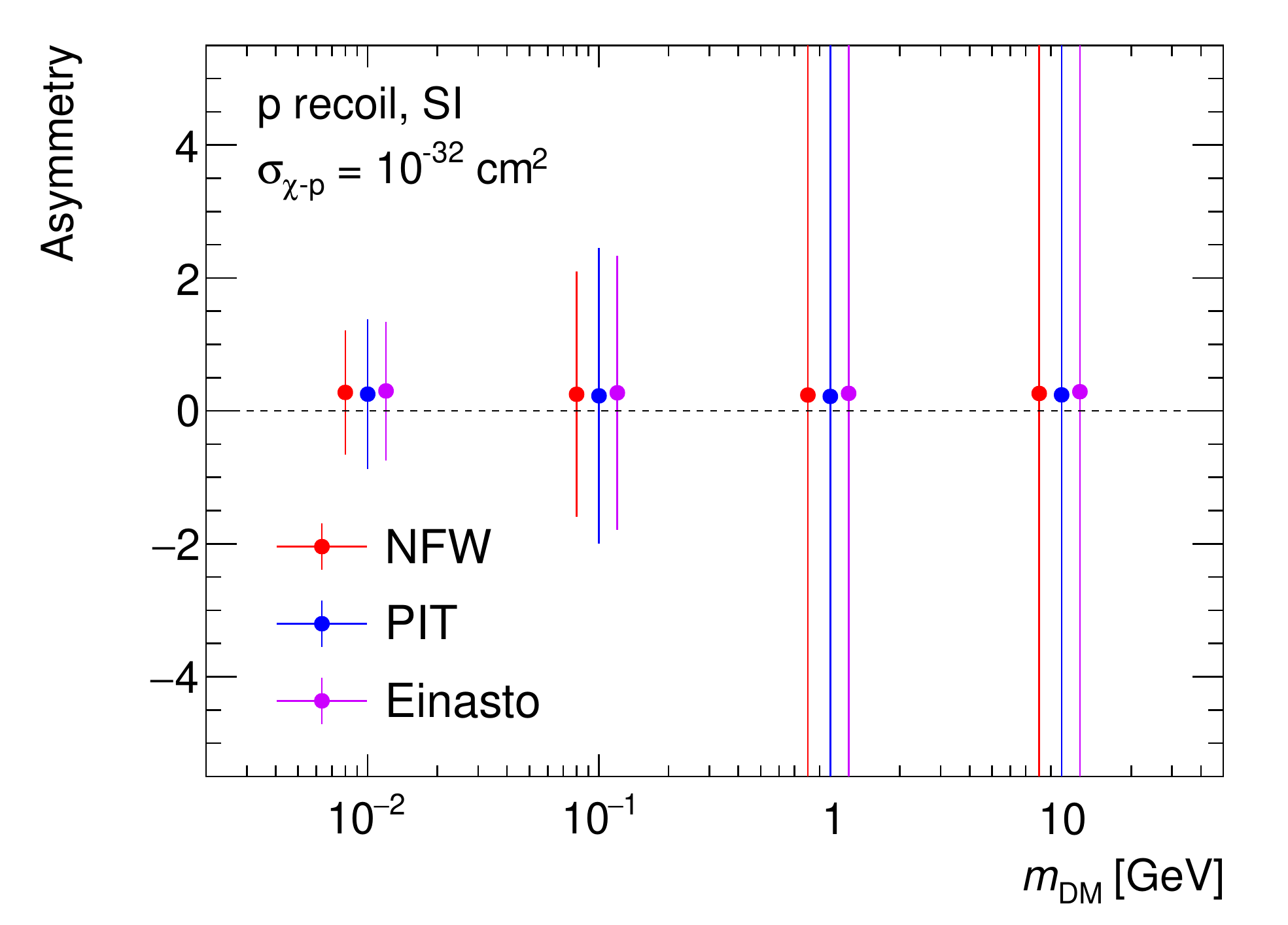}
    \vspace{-10mm}
    \caption*{$p$}
    \end{minipage}
    \begin{minipage}[b]{0.5\linewidth}
    \centering
    \includegraphics[keepaspectratio, width=\linewidth]{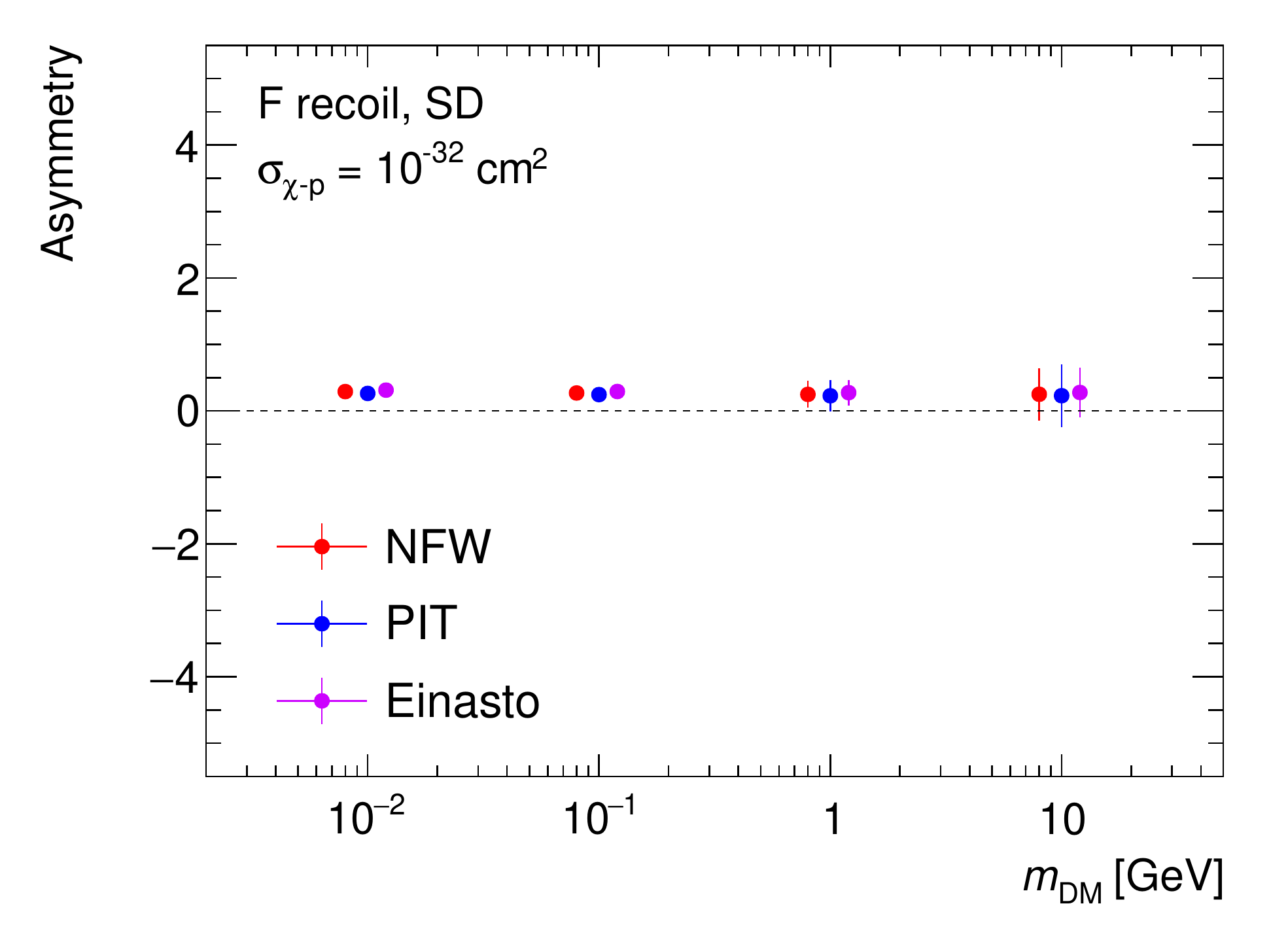}
    \vspace{-10mm}
    \caption*{F}
    \end{minipage}
 \end{tabular}
    \begin{tabular}{cc}
    \centering
    \begin{minipage}[b]{0.5\linewidth}
    \centering
    \includegraphics[keepaspectratio, width=\linewidth]{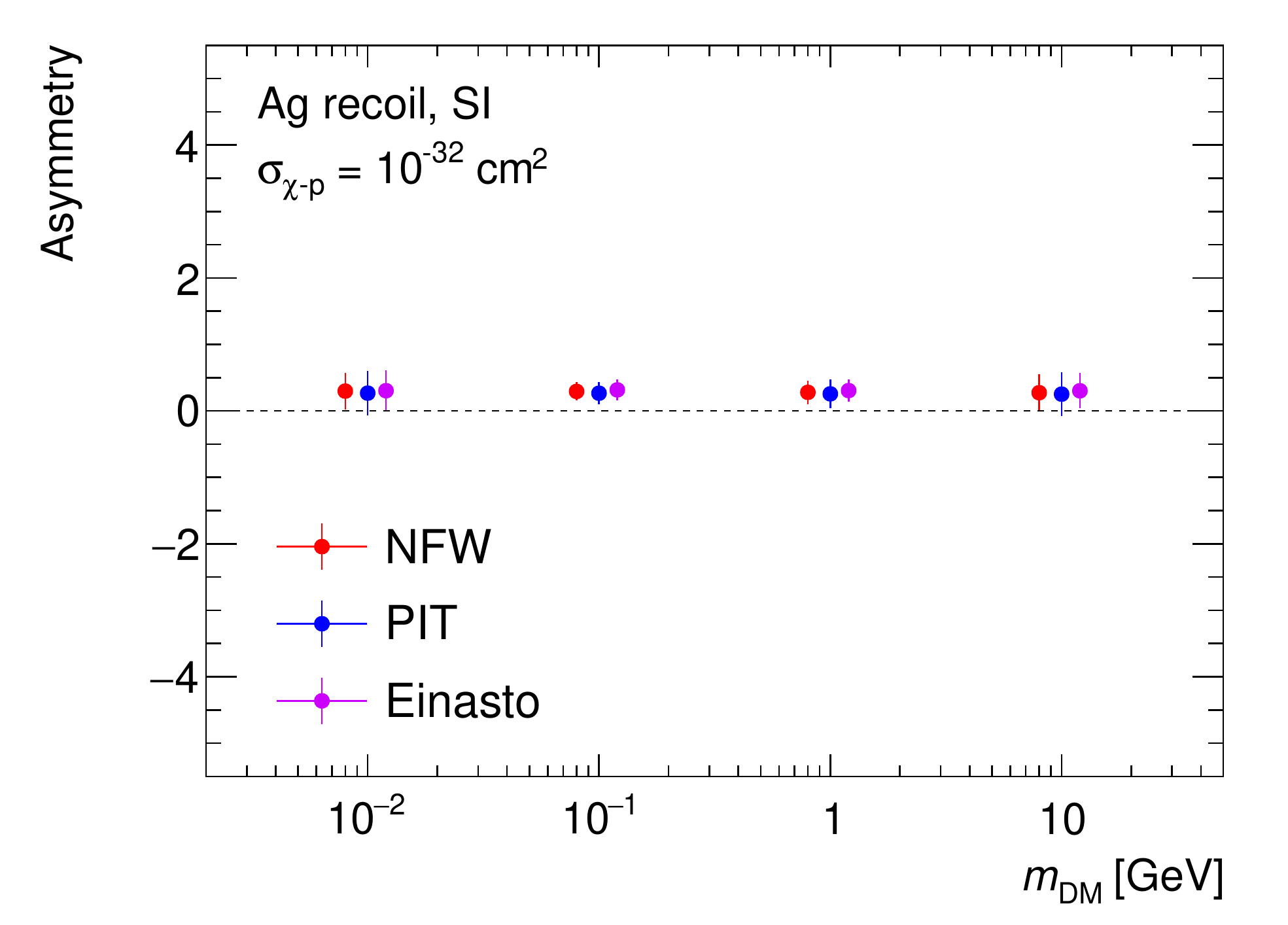}
    \vspace{-10mm}
    \caption*{Ag}
    \end{minipage}
 \end{tabular}
 \caption{Simulated asymmetry for each target and CR-DM mass. Colors indicate the density profiles. Scattering cross section for DM and proton is assumed to be $\sigma_{\chi p}=10^{-32}$~cm$^2$, and the exposure is assumed to be NIT 5.0~kg-yr for the target $p$ and Ag, and SF$_6$ 1550~kg-yr for the target F. }
 %Simulated asymmetry for each target and DM mass. Scattering cross section for DM and proton is assumed to be $\sigma_{\chi p}=10^{-32}$ cm$^2$, and the exposure is assumed to be NIT 2 kg-yr for the target p and Ag, and SF$_6$ 3.1 kg-yr for the target F.
 \label{Fig:asymm_mass}
\end{figure*}
%Asymmetry-DM mass plots end

%%%%%%%%%%%%%%%%%%%%
%Asymmetry plots
%%%%%%%%%%%%%%%%%%%%
\begin{figure*}[ht]
 \begin{tabular}{cc}
    \centering
\begin{minipage}[b]{0.5\linewidth}
    \centering
    \includegraphics[keepaspectratio, width=\linewidth]{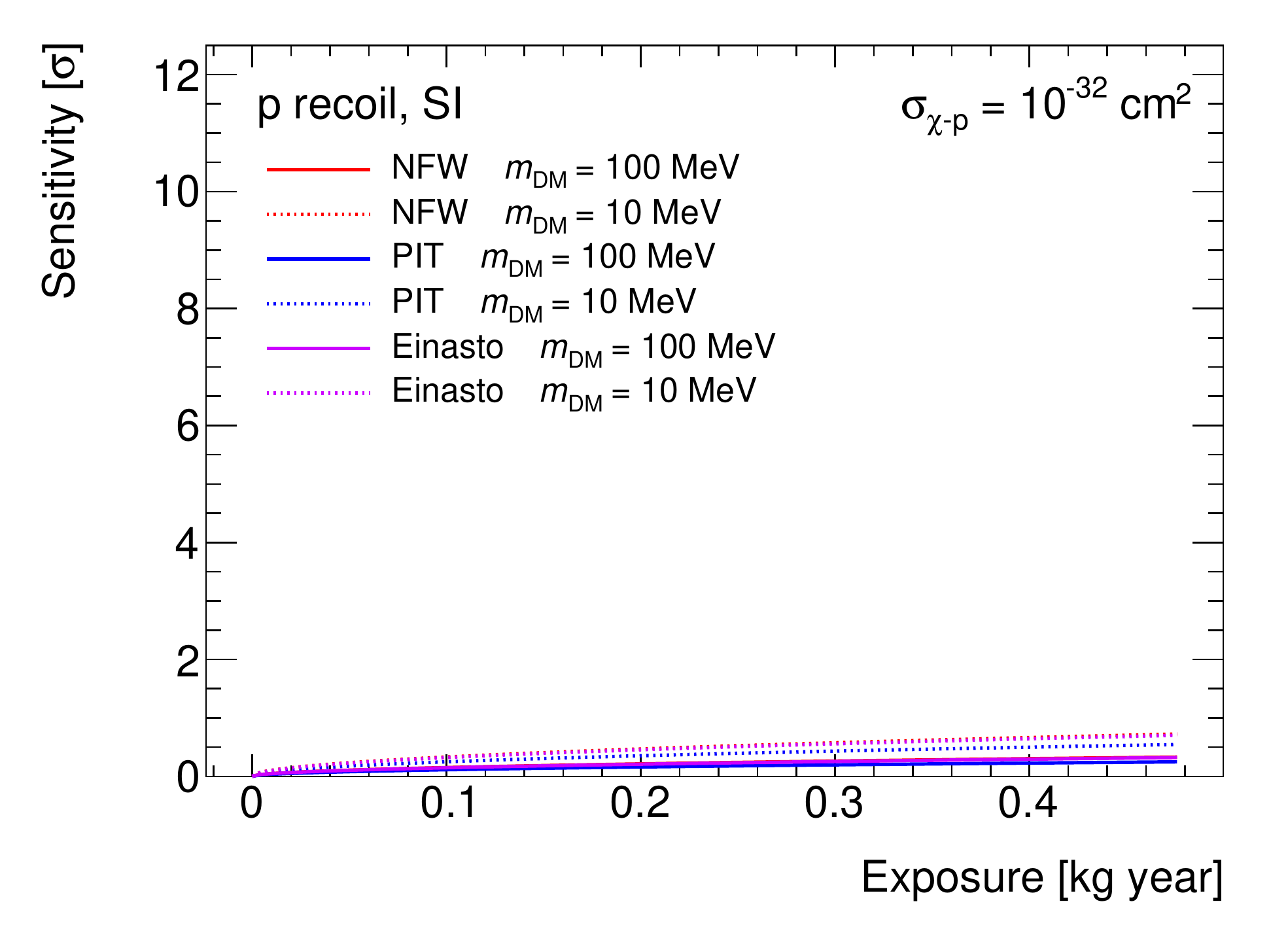}
    \vspace{-10mm}
    \caption*{$p$}
  \end{minipage}
  \begin{minipage}[b]{0.5\linewidth}
    \centering
    \includegraphics[keepaspectratio, width=\linewidth]{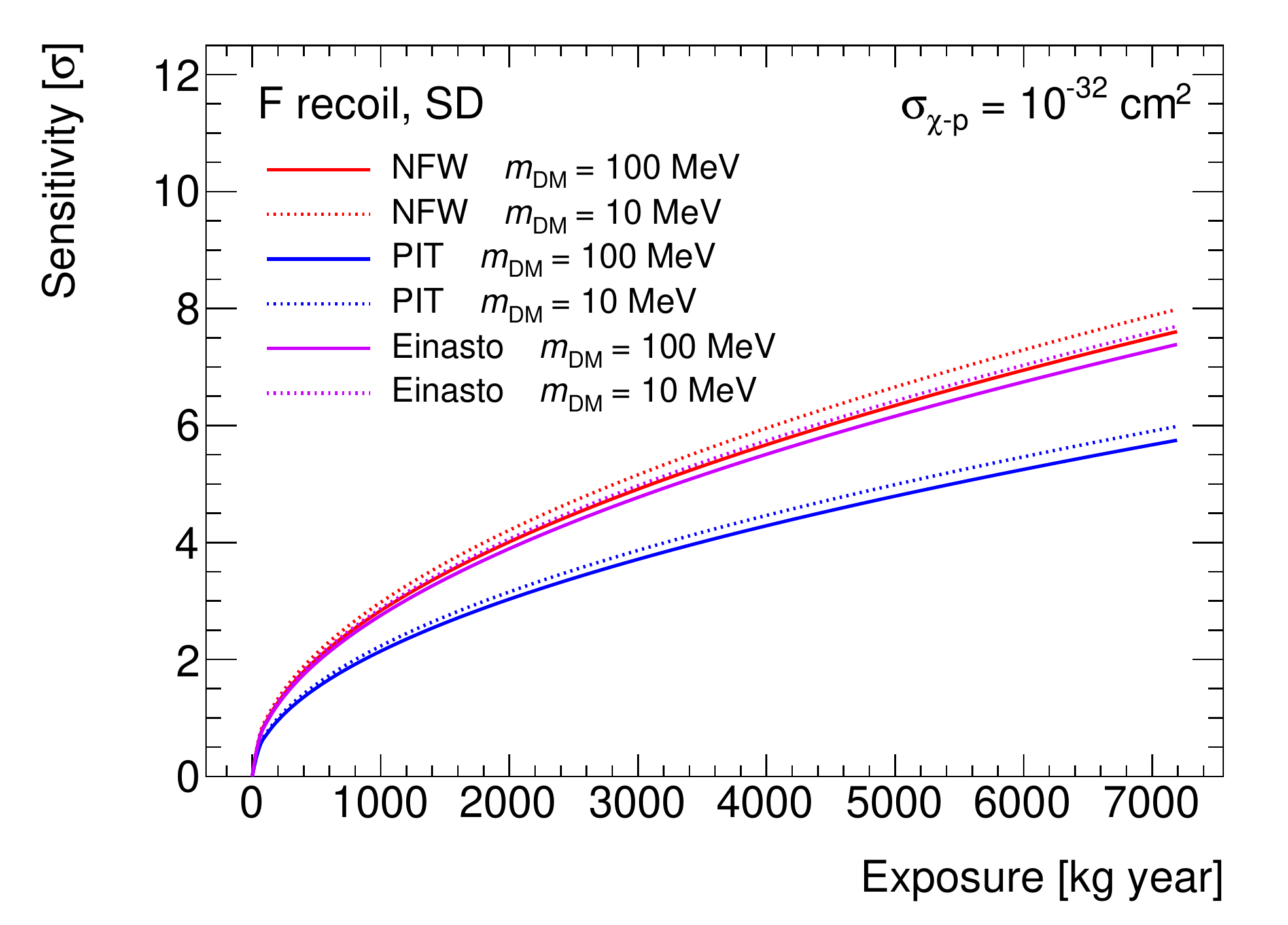}
    \vspace{-10mm}
    \caption*{F}
    \end{minipage}
 \end{tabular}
\begin{tabular}{cc}
    \centering
    \begin{minipage}[b]{0.5\linewidth}
    \centering
    \includegraphics[keepaspectratio, width=\linewidth]{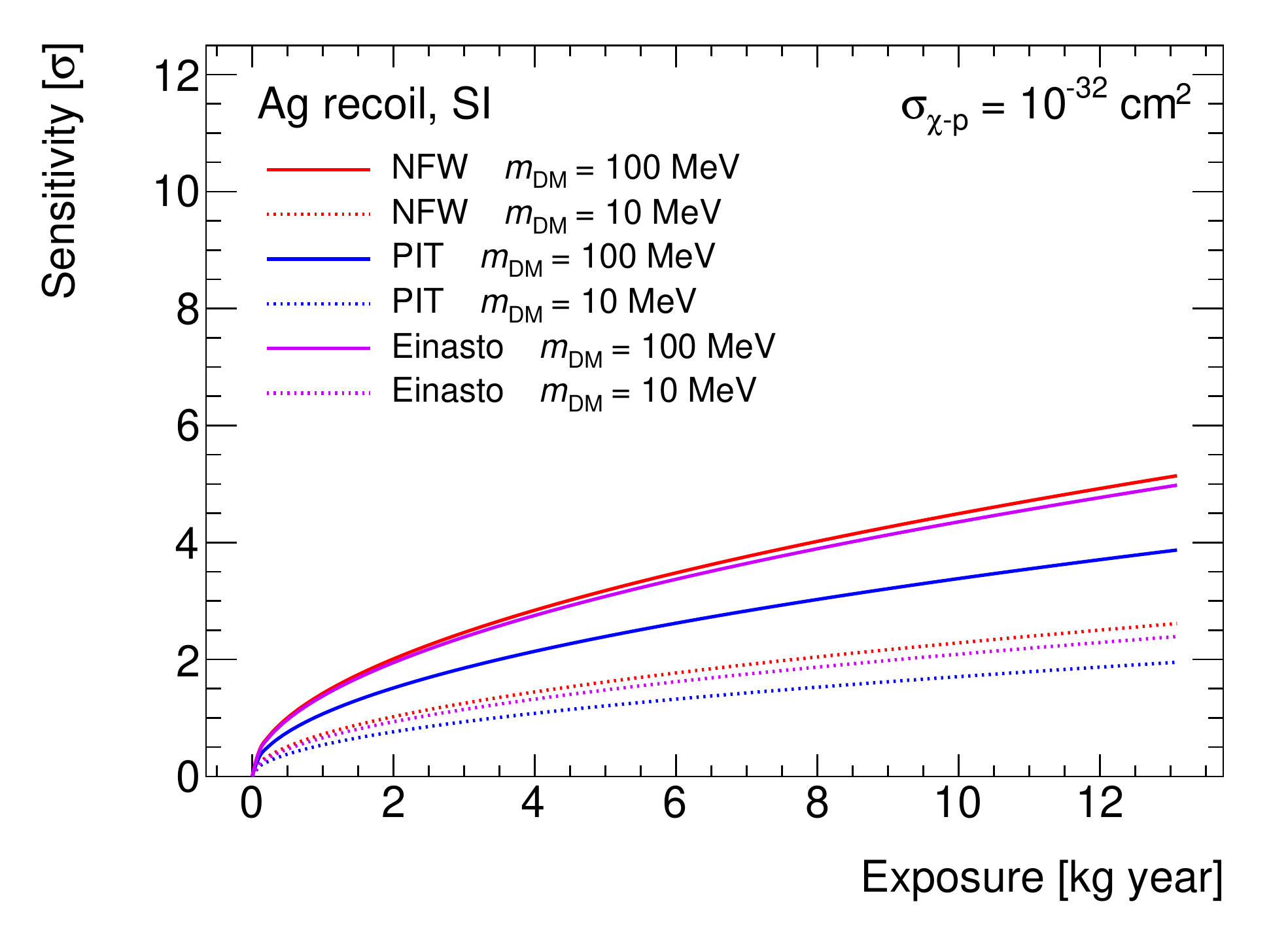}
    \vspace{-10mm}
    \caption*{Ag}
    \end{minipage}
 \end{tabular}
% \begin{tabular}{ccc}
%    \centering
%\begin{minipage}[b]{0.3\linewidth}
%    \centering
%    \includegraphics[keepaspectratio, width=\linewidth]{zcomp_p.pdf}
%    \caption*{$p$}
%  \end{minipage}
%  \begin{minipage}[b]{0.3\linewidth}
%    \centering
%    \includegraphics[keepaspectratio, width=\linewidth]{zcomp_F.pdf}
%    \caption*{F}
%    \end{minipage}
%    \begin{minipage}[b]{0.3\linewidth}
%    \centering
%    \includegraphics[keepaspectratio, width=\linewidth]{zcomp_Ag.pdf}
%    \caption*{Ag}
%    \end{minipage}
% \end{tabular}
 \caption{Null-rejection significance of the asymmetry parameter $A$ for each target. The CR-DM masses are assumed to be 10 MeV and 100~MeV, and are marked with dotted and solid lines, respectively. The scattering cross section for DM and proton is assumed to be $\sigma_{\chi p}=10^{-32}$~cm$^2$, which is the same as the preceding figures.}
 \label{Fig:z_target}
\end{figure*}
%Asymmetry plots end

To quantify the forward-backward asymmetries of the recoil direction distributions, a parameter $A$ is defined as
\begin{align}
A=\frac{n_{+}-n_{-}}{n_{+}+n_{-}},
\end{align}
%where $n_\pm$ 
where $n_+$ and $n_-$ are the number of events for $\cos\gamma_\mathrm{G.C.}>0$ and $\cos\gamma_\mathrm{G.C.}<0$, respectively. 

Figure~\ref{Fig:asymm_mass} shows the asymmetry parameters and their statistical errors obtained by the simulation. 
The colors represent the three density profiles and the errors are statistical ones.
%simply includes statistical uncertainty of $n_+$ and $n_-$.
%The error is simply estimated as $\delta A = \sqrt{(\partial A/\partial n_+)^2\delta n_+^2+(\partial A/\partial n_-)^2\delta n_-^2}$.
The number of events is normalized as 5.0~kg-yr exposure of the nano imaging tracker (NIT) for the target $p$ and Ag, and 1550~kg-yr exposure of SF$_6$ gas. 
The latter is based on the assumption of the exposure 10k~m$^3$ SF$_6$ gas at 20~Torr and six years, which is equivalent to 10 times the exposure envisioned for the Cygnus-1000 project ~\cite{Vahsen:2020pzb}.
The weight ratios of $p$ and Ag in the NIT are set at 1.6\% and 44.5\%, respectively~\cite{Asada:2017wvp}. 
%In the case of large CR-DM masses, the sensitivity to the asymmetry parameter is extremely low due to the lack of statistics.
%larger error. 
For light CR-DMs less than $O(10) - O(100)$~MeV mass, enough statistics clearly confirm the directionality originated by signals coming from the Galactic center direction. Because the number of events discarded for the light CR-DM instance, is smaller than for the heavy CR-DM instance and the central peak disappears.
As already mentioned, events with large recoil energy, which often occur in heavy CR-DM cases, are discarded because they are not elastic scattering, thus the number of events available for the analysis is reduced and the error is significant. 
The CR-DM generally can have a larger scattering cross section than WIMPs with masses around O(1 - 100) MeV. The large cross section often results in attenuation of flux before reaching underground detectors \cite{Bringmann:2018cvk, Ema:2018bih, Cappiello:2019qsw}. 
In the plots, we assume that the CR-DM descends from the upper atmosphere and reaches the underground detectors and neglect the attenuation effect. Due to their large scattering cross section, the component reaching the detector from the opposite side of the Earth experiences significant attenuation, thus it is not effective to use this component. Data selection is feasible for the time intervals when the detector is oriented towards the central region of the Galaxy, as gas detectors such as the NEWAGE offer time resolution, and in the NEWSdm, in principle it is possible to deactivate the nuclear emulsion by temperature adjustment \cite{Kimura:2017}. Thus, if such an event selection by time is performed, the net exposure would need to be doubled.
%The figures show that the Ag target has the smallest error and the highest sensitivity to the asymmetry parameter especially for keV- to MeV-scale CR-DM case. 
%In this case, where the mass difference between the target and the DM is large, the momentum transfer $T_R$ can be estimated to be
%\begin{align*}
%E_R^\mathrm{max}=\frac{m_\chi c^2}{ \sqrt{1-(v/c)^2}} \frac{4m_\chi m_N}{(m_\chi +m_N)^2}\simeq \frac{m_\chi c^2}{ \sqrt{1-(v/c)^2}} \frac{4m_\chi}{m_N},\\
%T_R&=\frac{T_\chi^2+2m_\chi T_\chi}{T_\chi + (m_\chi+m_N)^2/(2m_N)}\frac{1+\cos{\theta^*}}{2}\\
%&\simeq 4T_\chi\frac{m_\chi}{m_N} \frac{1+\cos{\theta^*}}{2}
%\end{align*}
%where $\theta^*$ is the scattering angle in the center-of-mass system and $T_\chi$ is the kinetic energy of the CR-DM which is as small as MeV-scale for light DM with MeV-scale mass~\cite{Bringmann:2018cvk}.
%The momentum transfer is highly suppressed for light DM, resulting in many low-energy recoil events available for the analysis.

%In Figure~\ref{table:FOM}, the total number of events $n_\mathrm{tot}$, $n_+$, the asymmetry $A$, and its error $\delta A$ are shown for each case, where $n_\mathrm{tot}$ is of course $n_\mathrm{tot}=n_++n_-$.
%The event numbers $n_\mathrm{tot}$ and $n_+$ are normalized as 0.1-kg-yr exposure of the nuclear emulsion for p and Ag, and 1-kg-yr exposure of SF$_6$ gas for F. 
%Those errors are simply estimated as $\delta n_\pm=\sqrt{n_\pm},\  \delta n_\mathrm{tot}=\sqrt{(\partial n_\mathrm{tot}/\partial n_+)^2\delta n_+^2+(\partial n_\mathrm{tot}/\partial n_-)^2\delta n_-^2}$.

Figure~\ref{Fig:z_target} shows the evolution of the null-rejection significance of the asymmetry parameter $A$ with each target, where the range of the exposure corresponds up to six years for the same configuration as that in Figure~\ref{Fig:asymm_mass}. 
%The scattering cross sections for DM and protons $\sigma_{\chi-p}$ is assumed to be 10$^{-32}$ cm$^2$ in each case, and 
The CR-DM masses are assumed to be 10~MeV (dotted lines) and 100~MeV (solid lines). 
As can be inferred from Figure~\ref{Fig:asymm_mass}, assuming elastic scattering, %the target Ag is the most sensitive to the asymmetry parameter.
all targets investigated have sufficient potential for the null-rejection of the asymmetry parameter $A$ within the scope of the envisioned future upgrades.
%In the figure, exposures are assumed to be higher than those used in the current plan, i.e., the exposures on the right side of the figures correspond to five times of the currently assumed exposures. 
With enough exposure, the asymmetry can be confirmed with more than 5~$\sigma$ confidence level for 10~MeV (100~MeV) mass CR-DM except for $p$, at most  8~$\sigma$ (7.6~$\sigma$) and 2.6~$\sigma$ (5.2~$\sigma$) for F and Ag, respectively.
For SD interactions using the fluorine target, the NEWAGE experiment is developing a 1~m$^{3}$ scale detector filled with SF$_{6}$ gas. Furthermore, an update is planned to the CYGNUS-1000 experiment~\cite{Vahsen:2020pzb}, which has 1000~m$^{3}$ scale gaseous time projection chamber (TPC). It is planned to achieve 5~$\sigma$ sensitivity for the 10~MeV CR-DM mass after 2.6 (2.4) ~years' of measurement for Einasto (NFW) profile.
For the NEWSdm experiment, the device production system at the Gran Sasso National Laboratory (LNGS) has been operated successfully, and it is possible to produce the 10~kg scale target using the current system. In addition, new scanning systems with a 0.5~kg/year/machine are being developed, and several kg-scale scanning will be achieved by using the same upgrades on five existing machines. Furthermore, a high scanning speed machine with a $>$ 5~kg/year/machine has already been designed that uses a wide field of view with multi-camera imaging and optimal driving motion system. From the high scanning machine, several 10~kg scale experiment will be achieved. Low energy threshold tracking will be achieved by the finer grain nuclear emulsions~\cite{Asada:2017wvp} and super-resolution techniques~\cite{Andrey:2020}.
Therefore, taking into account the possibility of future upgrades, both NEWAGE and NEWSdm would have the potential to verify the light CR-DM arriving from the direction of the Galactic center.

\section{Conclusion}
\label{sec:Conclusion}
Up-scattering of DM by cosmic rays is an undeniable key phenomenon that can allow the search for 
%verification of 
light DMs, which are not easy to reach by standard nuclear recoil methods. The CR-DMs are expected to arrive mostly from the direction of the Galactic center, where the density of the DM is large, and thus can be 
%reasonably 
examined with direction-sensitive detectors. In this paper, we study the detection possibilities of CR-DMs with directional detectors and investigate the expected energy and angular distributions.

We find that almost independent of the DM density profiles, CR-DM tends to come from the direction of the Galactic center, and the targets in the detector are scattered towards the forward direction.
%reflecting the direction. The angular distributions of the nuclear recoils show clear peaks at $\cos{\gamma_\mathrm{G.C.}}=1$ for the CR-DM with masses of keV- to sub-GeV scales as a consequence of the forward scatterings.
%The CR-DMs with masses of the keV-scale%, where elastic scattering accounts for most of the scattering, %(?), 
Assuming elastic scattering of DM and target, we also investigate the sensitivity of directional detectors to verify the existence of CR-DM.
Our study showed that the asymmetry of nuclear recoil is expected to hold the characteristic signature of the CR-DM. 
It is shown that the feature can be verified by directional detectors, especially for light DM with mass less than $O(100)-O(10)$~MeV.
With scattering cross section of DM and proton as $\sigma_{\chi-p}=10^{-32}$~cm$^2$ and $m_\chi=10$~MeV ($m_\chi=100$~MeV), the asymmetry can be evaluated at 0.7$\sigma$, 8$\sigma$, and 2.6$\sigma$ (0.2$\sigma$, 7.6$\sigma$, and 5.2$\sigma$) at most for the $p$ target (the SI interaction), for the F target (the SD interaction), and for the Ag target (the SI interaction), respectively. For the evaluation, exposures 0.48~kg-yr, 7200~kg-yr, and 13.2~kg-yr are 
assumed for $p$, F, and Ag targets, respectively.
%Although NEWSdm's and NEWAGE's current exposure is about 1/100th of those, 
These large exposures can be achieved with future large detectors like CYGNUS and NEWSdm. 
%y are beyond the scope of NEWSdm's and NEWAGE's current exposure, however, }
%they have future prospects for upgrades to reach enough sensitivity to investigate the directional asymmetry of incoming CR-DM.

\clearpage
%%%%%%%%%%%%%%%%%%%%
% sky map plots -DM-
%%%%%%%%%%%%%%%%%%%%
\setcounter{figure}{1}
\begin{figure*}[ht]
 \begin{tabular}{ccc}
    \centering
    \begin{minipage}[b]{0.3\linewidth}
    \centering
    \includegraphics[keepaspectratio, width=\linewidth]{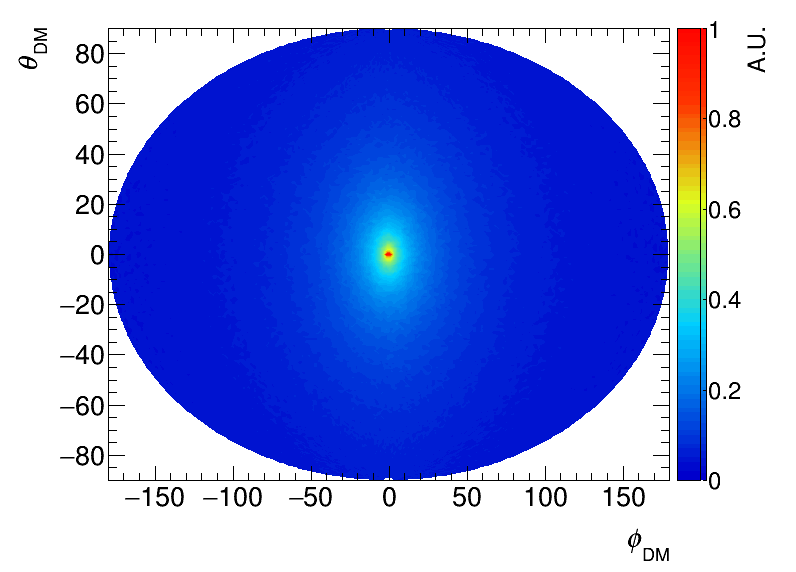}
    \caption*{NFW}
  \end{minipage}
  \begin{minipage}[b]{0.3\linewidth}
    \centering
    \includegraphics[keepaspectratio, width=\linewidth]{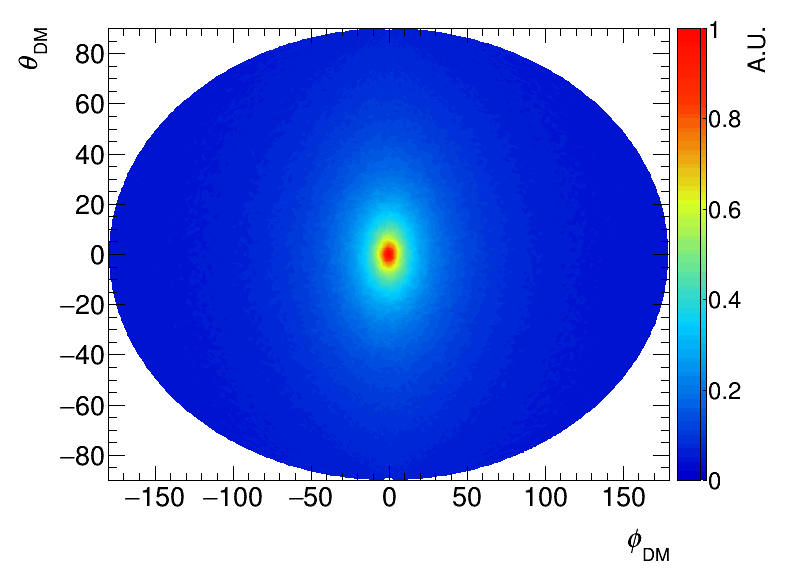}
    \caption*{Einasto}
    \end{minipage}
    \begin{minipage}[b]{0.3\linewidth}
    \centering
    \includegraphics[keepaspectratio, width=\linewidth]{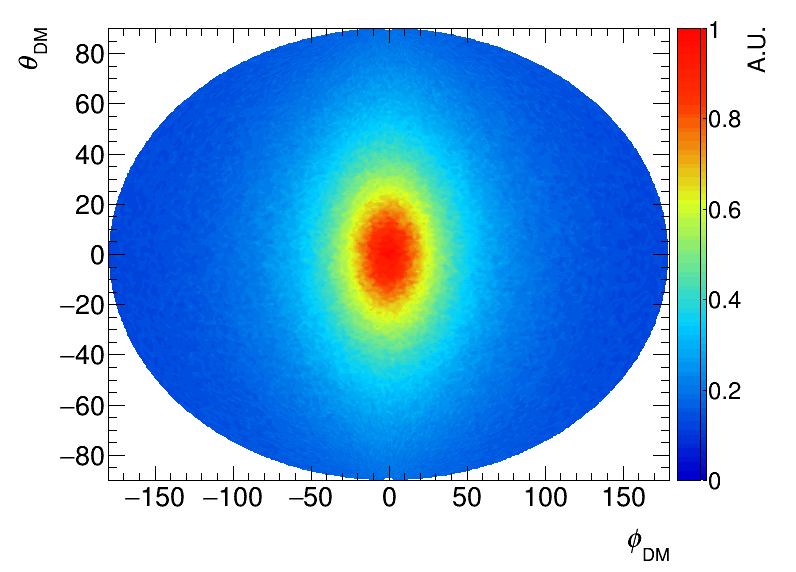}
    \caption*{PIT}
    \end{minipage}
 \end{tabular}
 \caption{Sky map of the CR-DM in the Solar system. The left-handed side, the center, and the right-handed side correspond to the NFW, the Einasto, and the PIT profiles, respectively. The CR-DM mass is 10 MeV. The density of the number of events is written using arbitrary units (A.U.). }%Note that angle $\phi_\mathrm{DM}$ is defined as $\phi_\mathrm{DM}=\phi+\pi$ where $\phi$ is in Figure~\ref{fig:coordinatesystem.pdf}.}}
 \label{Fig:skymap_dm}
\end{figure*}
%%%%%%%%%%%%%%%%%%%%%%%%%%%%%%%%%%%%%%%%%%%%%%%%%%%%%%%%%%%%%%%%%%%%%%%%%%%%%%%%%

%%%%%%%%%%%%%%%%%%%%
% sky map plots -recoil-
%%%%%%%%%%%%%%%%%%%%
\begin{figure*}[t]
 \begin{tabular}{ccc}
    \centering
    \begin{minipage}[b]{0.3\linewidth}
    \centering
    \includegraphics[keepaspectratio, width=\linewidth]{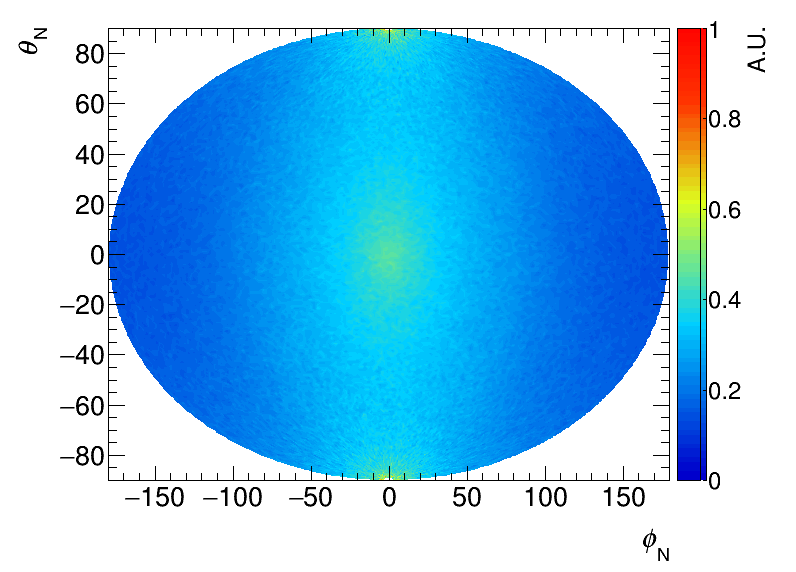}
    \caption*{NFW, $m_\chi=100$ MeV}
  \end{minipage}
  \begin{minipage}[b]{0.3\linewidth}
    \centering
    \includegraphics[keepaspectratio, width=\linewidth]{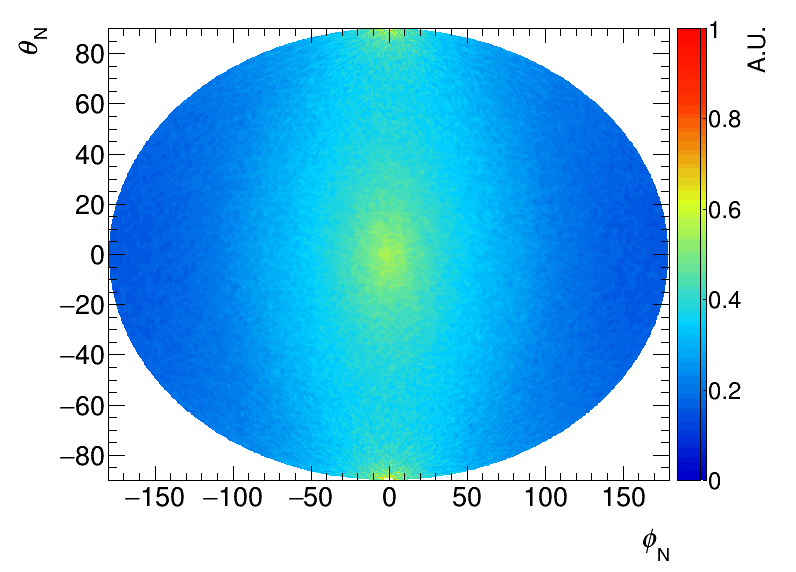}
    \caption*{Einasto, $m_\chi=100$ MeV}
    \end{minipage}
    \begin{minipage}[b]{0.3\linewidth}
    \centering
    \includegraphics[keepaspectratio, width=\linewidth]{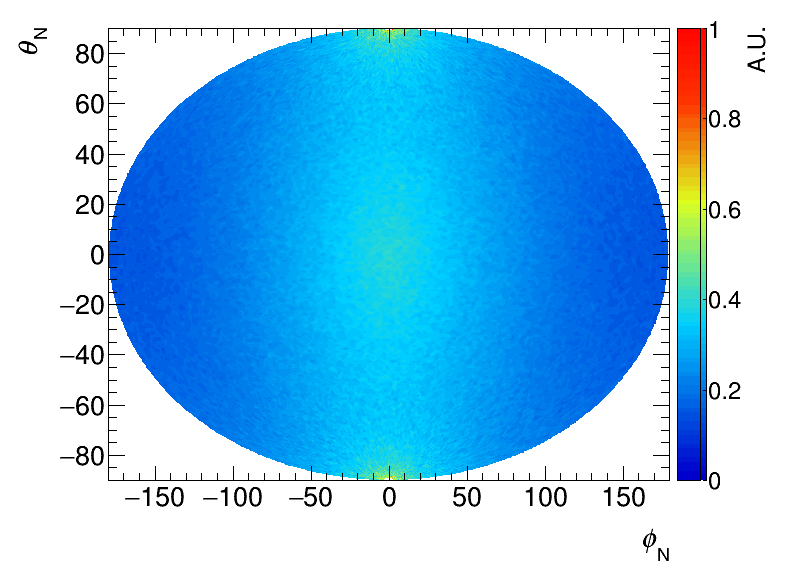}
    \caption*{PIT,  $m_\chi=100$ MeV}
    \end{minipage}
 \end{tabular}
 \begin{tabular}{ccc}
    \centering
    \begin{minipage}[b]{0.3\linewidth}
    \includegraphics[keepaspectratio,width=\linewidth]{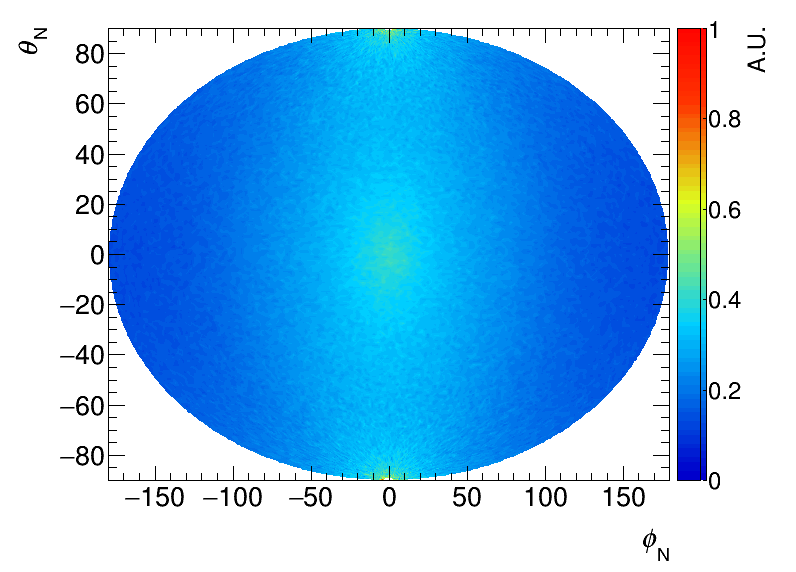}
    \caption*{NFW, $m_\chi=10$ MeV}
  \end{minipage}
  \begin{minipage}[b]{0.3\linewidth}
    \centering
    \includegraphics[keepaspectratio, width=\linewidth]{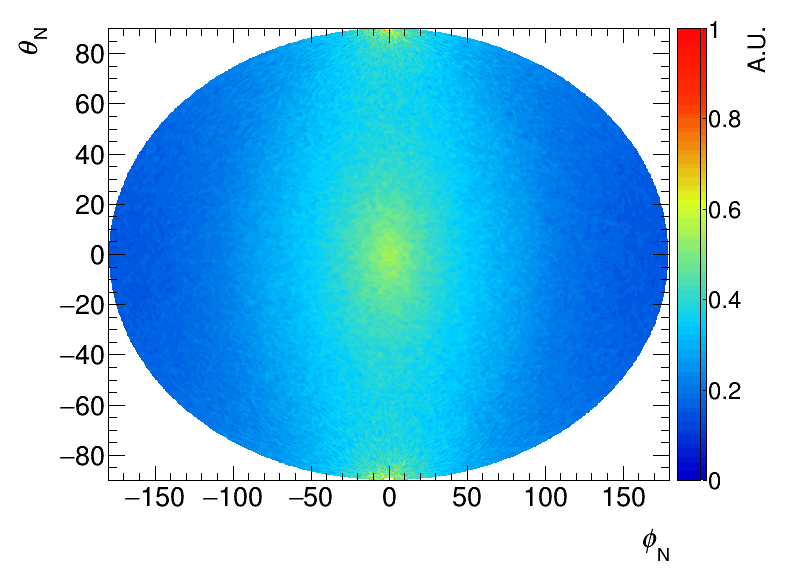}
    \caption*{Einasto, $m_\chi=10$ MeV}
    \end{minipage}
    \begin{minipage}[b]{0.3\linewidth}
    \centering
    \includegraphics[keepaspectratio, width=\linewidth]{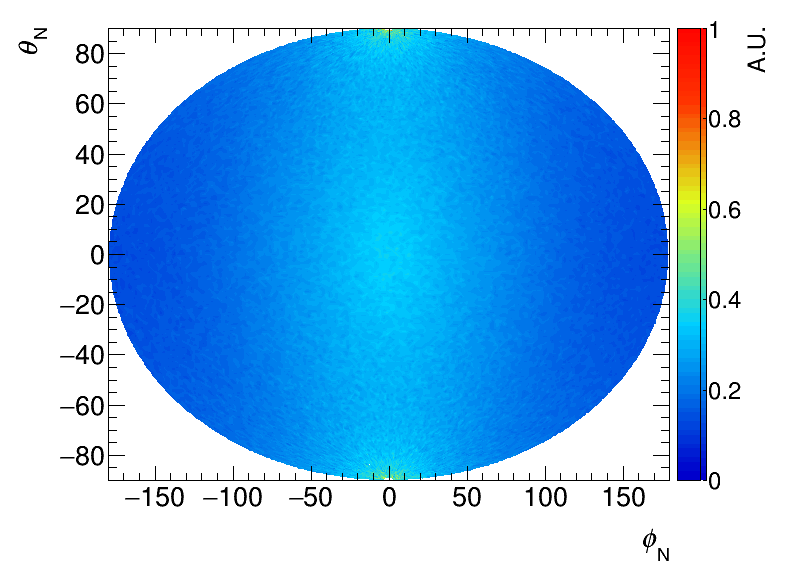}
    \caption*{PIT, $m_\chi=10$ MeV}
    \end{minipage}
 \end{tabular}
 \begin{tabular}{ccc}
    \centering
    \begin{minipage}[b]{0.3\linewidth}
    \centering
    \includegraphics[keepaspectratio, width=\linewidth]{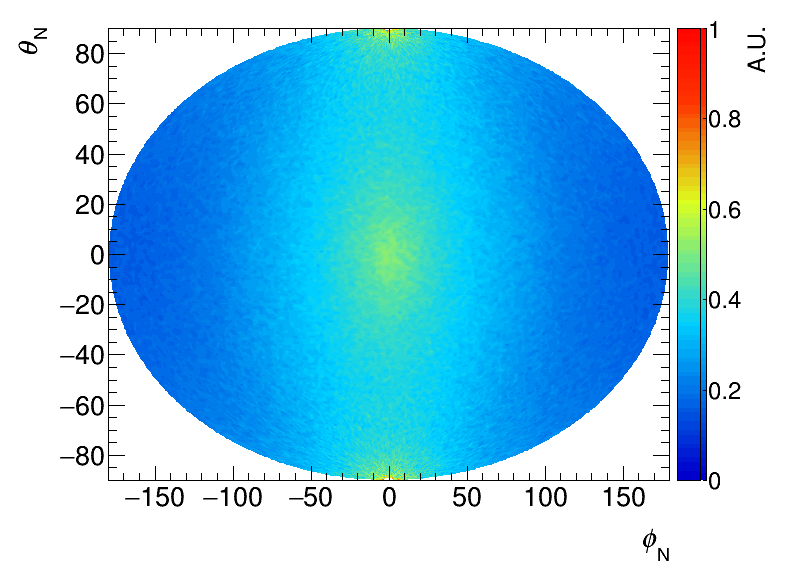}
    \caption*{NFW, $m_\chi=1$ MeV}
  \end{minipage}
  \begin{minipage}[b]{0.3\linewidth}
    \centering
    \includegraphics[keepaspectratio, width=\linewidth]{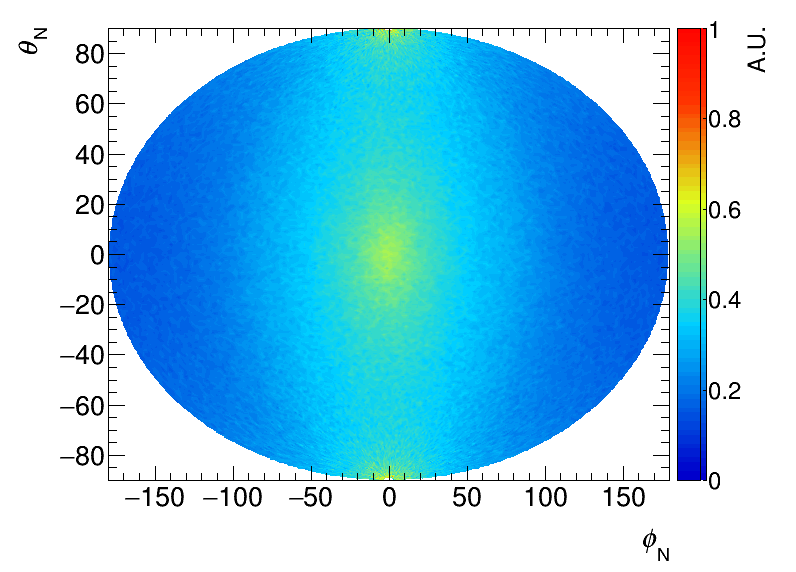}
    \caption*{Einasto, $m_\chi=1$ MeV}
    \end{minipage}
    \begin{minipage}[b]{0.3\linewidth}
    \centering
    \includegraphics[keepaspectratio, width=\linewidth]{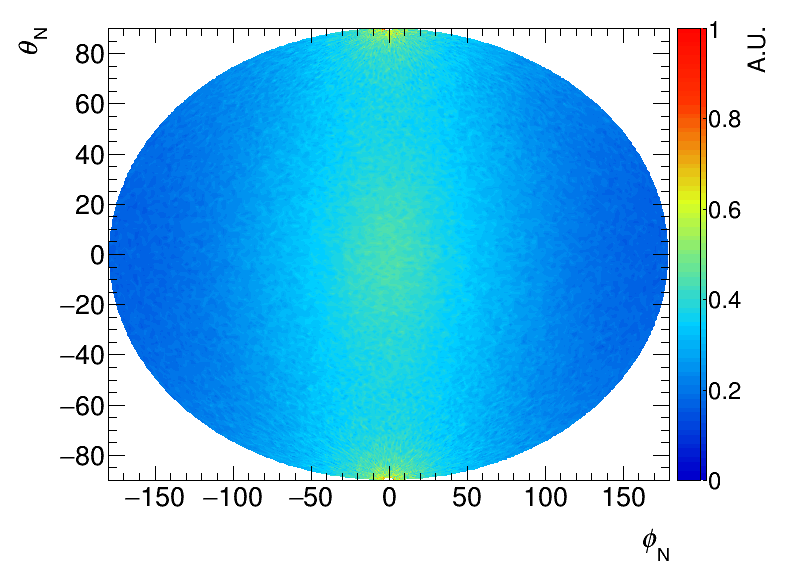}
    \caption*{PIT, $m_\chi=1$ MeV}
    \end{minipage}
 \end{tabular}
 \caption{Sky map of the nuclear recoil directions in the Solar system. The left column, the center column, and the right column correspond to the NFW, the Einasto, and the PIT profiles, respectively. The upper row corresponds to the CR-DM mass of 100~MeV, the middle row to 10~MeV, and the lower row to 1~MeV. The nuclear target is $p$. We assume SI interactions between the DM particle and the nuclear target.}
 \label{Fig:skymap_p}
\end{figure*}
%%%%%%%%%%%%%%%%%%%%%%%%%%%%%%%%%%%%%%%%%%%%%%%%%%%%%%%%%%%%%%%%%%%%%%%%%%%%%%%%%
\begin{figure*}[t]
 \begin{tabular}{ccc}
    \centering
    \begin{minipage}[b]{0.3\linewidth}
    \centering
    \includegraphics[keepaspectratio, width=\linewidth]{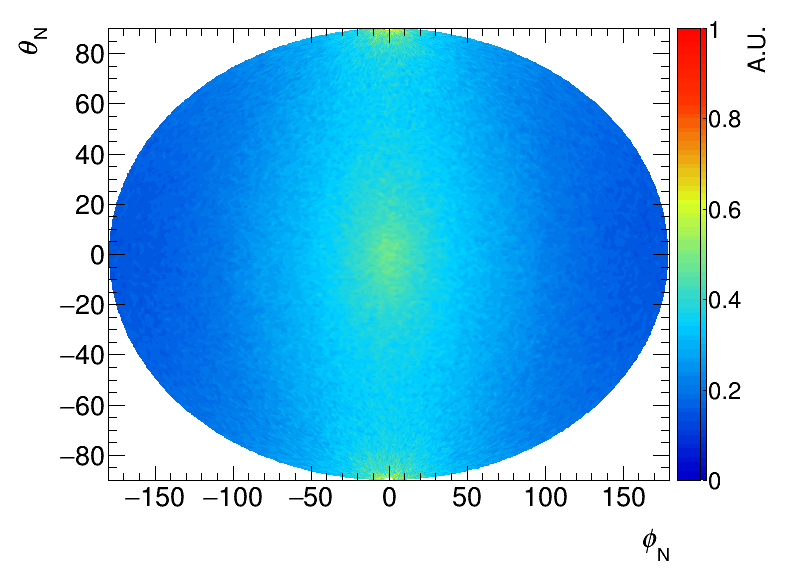}
    \caption*{NFW, $m_\chi=100$ MeV}
  \end{minipage}
  \begin{minipage}[b]{0.3\linewidth}
    \centering
    \includegraphics[keepaspectratio, width=\linewidth]{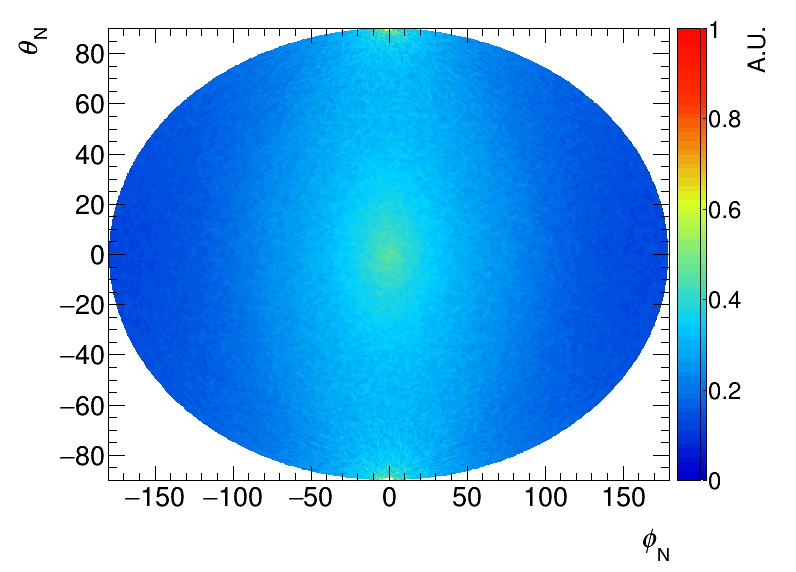}
    \caption*{Einasto, $m_\chi=100$ MeV}
    \end{minipage}
    \begin{minipage}[b]{0.3\linewidth}
    \centering
    \includegraphics[keepaspectratio, width=\linewidth]{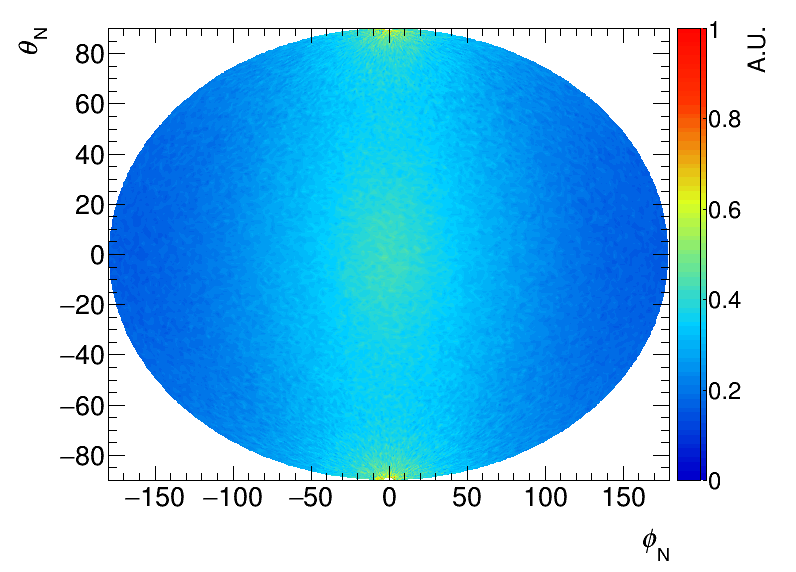}
    \caption*{PIT, $m_\chi=100$ MeV}
    \end{minipage}
 \end{tabular}
 \begin{tabular}{ccc}
    \begin{minipage}[b]{0.3\linewidth}
    \centering
    \includegraphics[keepaspectratio, width=\linewidth]{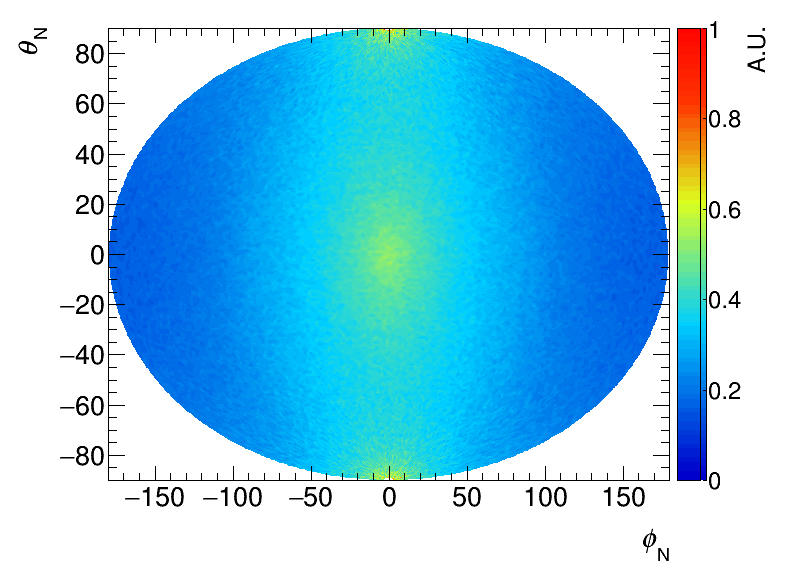}
    \caption*{NFW, $m_\chi=10$ MeV}
     \end{minipage}
  \begin{minipage}[b]{0.3\linewidth}
    \centering
    \includegraphics[keepaspectratio, width=\linewidth]{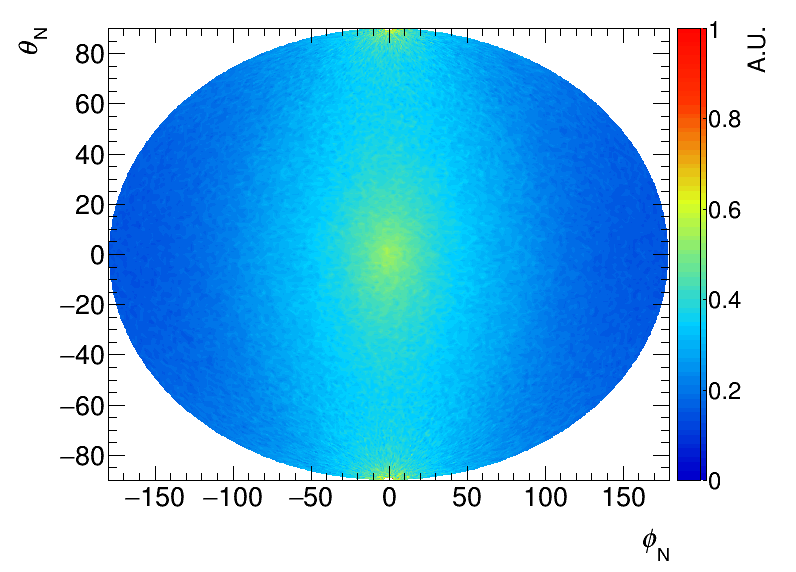}
    \caption*{Einasto, $m_\chi=10$ MeV}
    \end{minipage}
    \begin{minipage}[b]{0.3\linewidth}
    \centering
    \includegraphics[keepaspectratio, width=\linewidth]{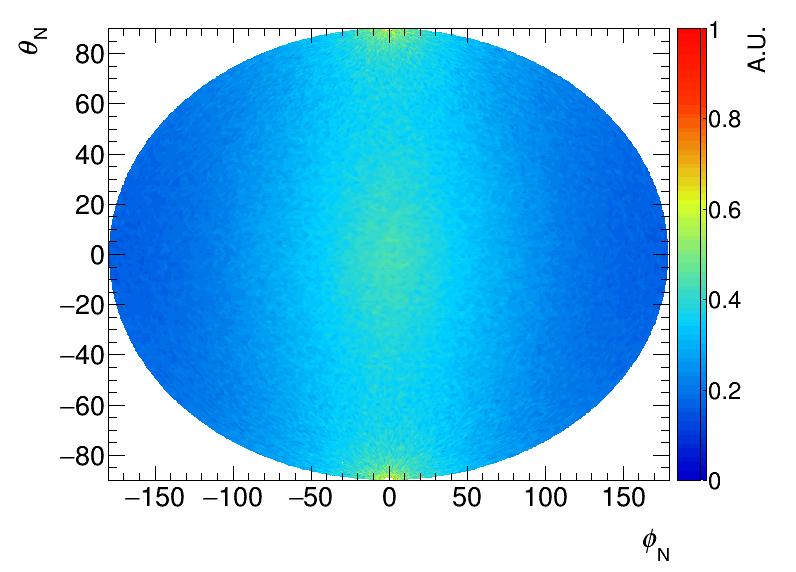}
    \caption*{PIT, $m_\chi=10$ MeV}
    \end{minipage}
 \end{tabular}
 \begin{tabular}{ccc}
    \centering
    \begin{minipage}[b]{0.3\linewidth}
    \centering
    \includegraphics[keepaspectratio, width=\linewidth]{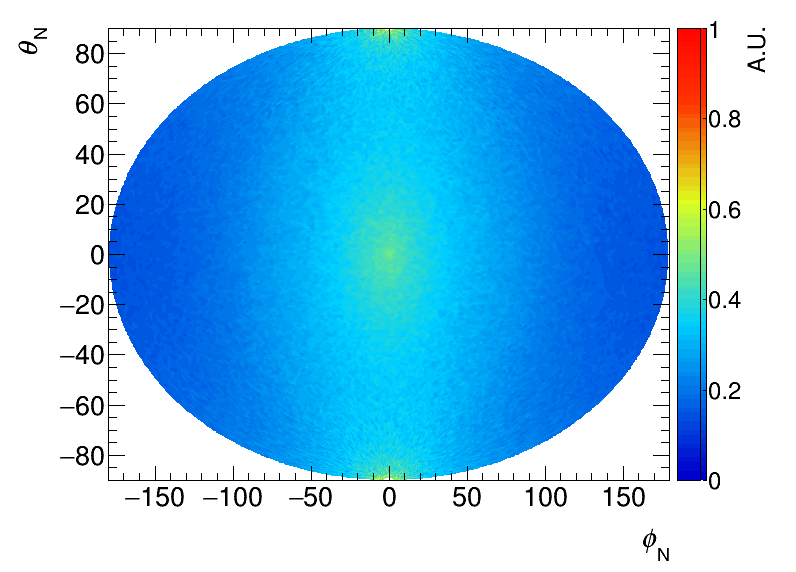}
    \caption*{NFW, $m_\chi=1$ MeV}
  \end{minipage}
  \begin{minipage}[b]{0.3\linewidth}
    \centering
    \includegraphics[keepaspectratio, width=\linewidth]{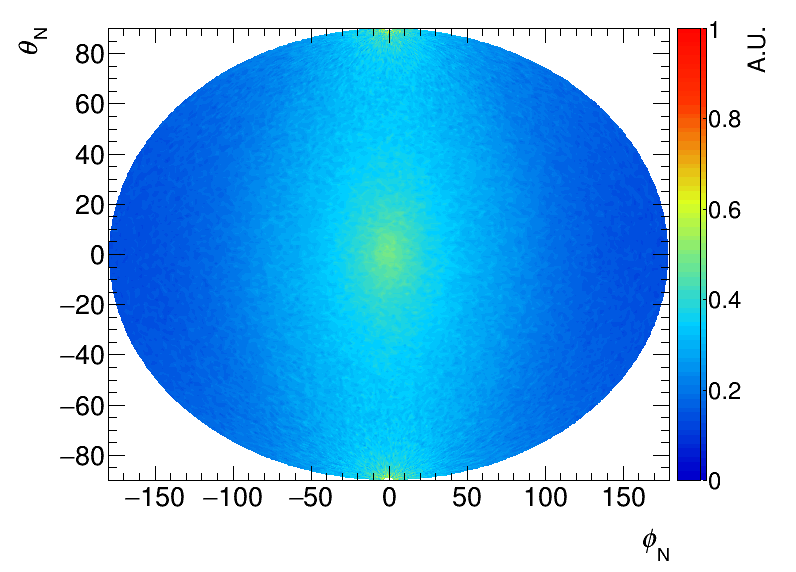}
    \caption*{Einasto, $m_\chi=1$ MeV}
    \end{minipage}
    \begin{minipage}[b]{0.3\linewidth}
    \centering
    \includegraphics[keepaspectratio, width=\linewidth]{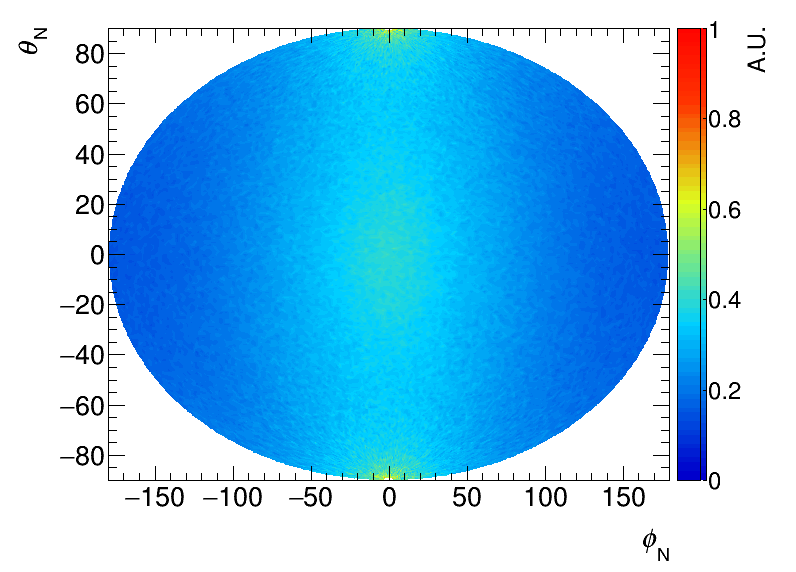}
    \caption*{PIT, $m_\chi=1$ MeV}
    \end{minipage}
 \end{tabular}
 \caption{Legend is same as Figure~\ref{Fig:skymap_p} except that the target atom in the direct detection is F.}
 \label{Fig:skymap_F}
\end{figure*}
%%%%%%%%%%%%%%%%%%%%%%%%%%%%%%%%%%%%%%%%%%%%%%%%%%%%%%%%%%%%%%%%%%%%%%%%%%%%%%%%%
\begin{figure*}[t]
  \begin{tabular}{ccc}
    \centering
    \begin{minipage}[b]{0.3\linewidth}
    \centering
    \includegraphics[keepaspectratio, width=\linewidth]{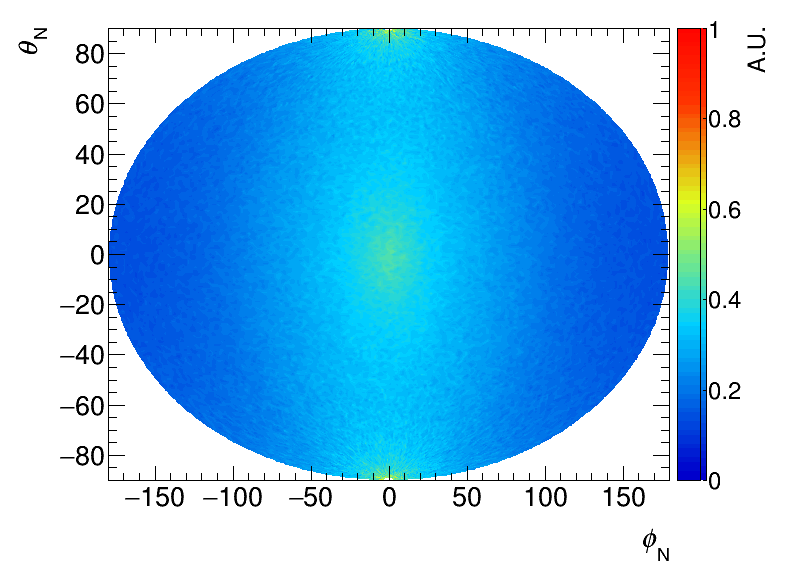}
    \caption*{NFW, $m_\chi=100$ MeV}
  \end{minipage}
  \begin{minipage}[b]{0.3\linewidth}
    \centering
    \includegraphics[keepaspectratio, width=\linewidth]{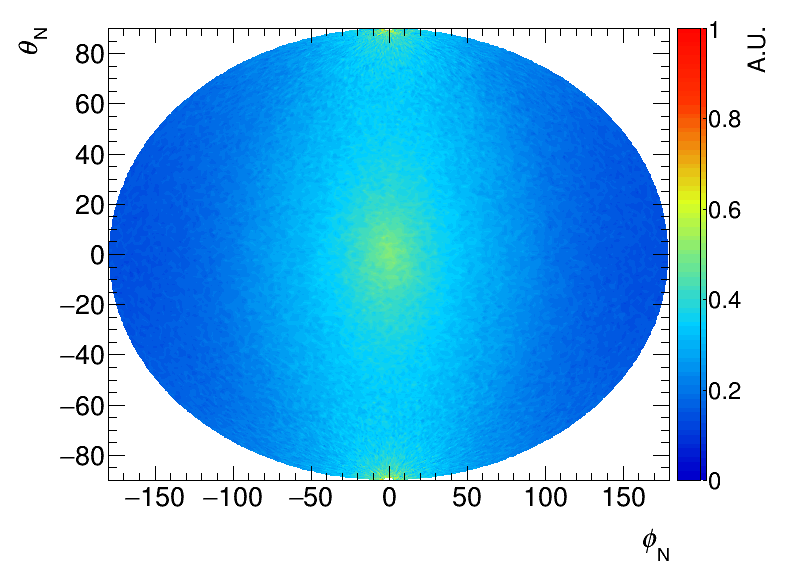}
    \caption*{Einasto, $m_\chi=100$ MeV}
    \end{minipage}
    \begin{minipage}[b]{0.3\linewidth}
    \centering
    \includegraphics[keepaspectratio, width=\linewidth]{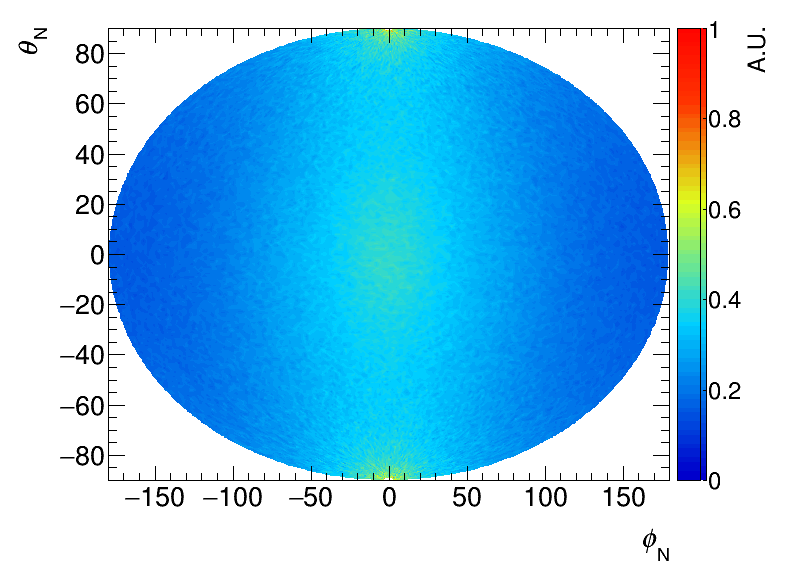}
    \caption*{PIT, $m_\chi=100$ MeV}
    \end{minipage}
 \end{tabular}
 \begin{tabular}{ccc}
    \centering
    \begin{minipage}[b]{0.3\linewidth}
    \includegraphics[keepaspectratio, width=\linewidth]{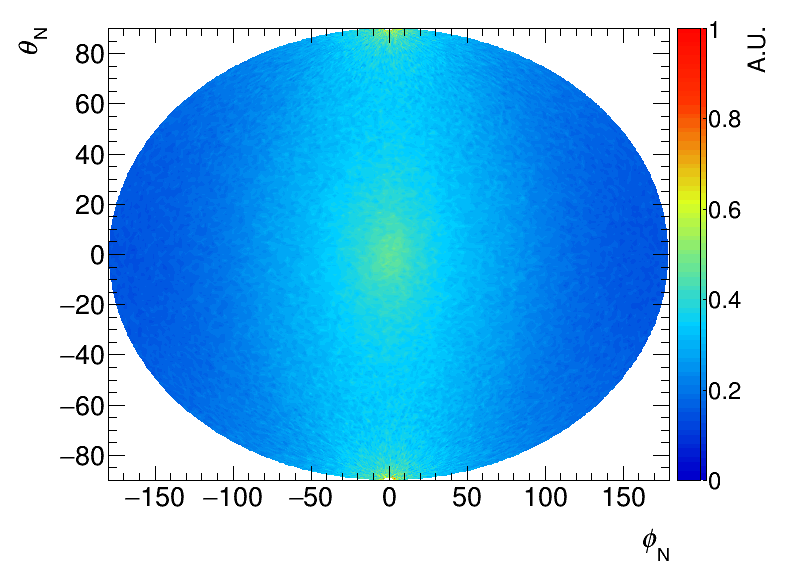}
    \caption*{NFW, $m_\chi=10$ MeV}
  \end{minipage}
  \begin{minipage}[b]{0.3\linewidth}
    \centering
    \includegraphics[keepaspectratio, width=\linewidth]{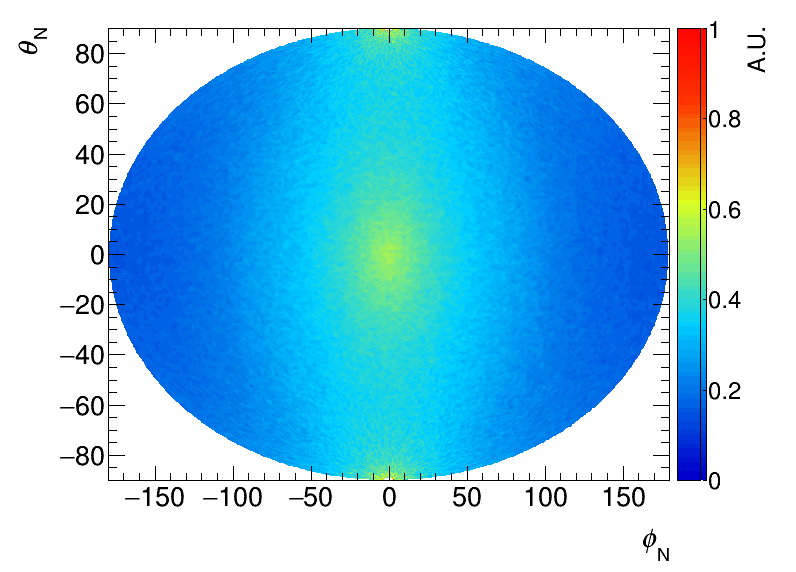}
    \caption*{Einasto, {$m_\chi=10$ MeV}}
    \end{minipage}
    \begin{minipage}[b]{0.3\linewidth}
    \centering
    \includegraphics[keepaspectratio, width=\linewidth]{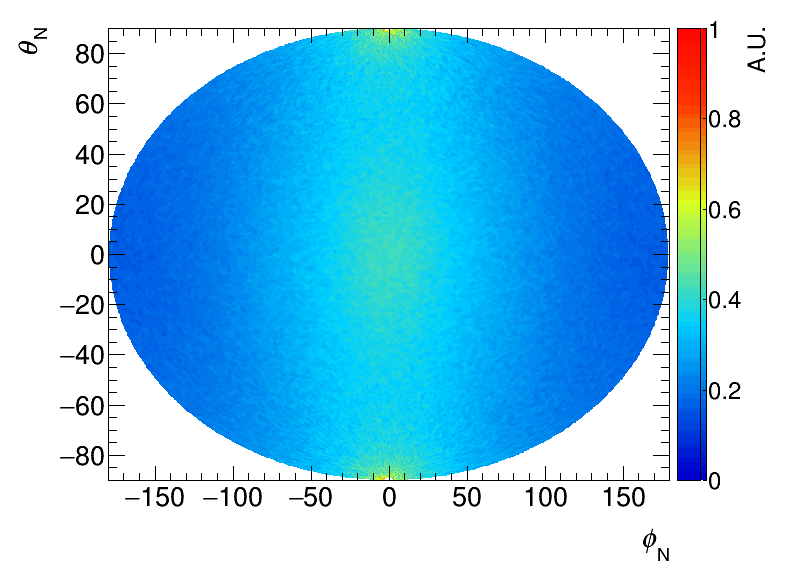}
    \caption*{PIT, $m_\chi=10$ MeV}
    \end{minipage}
 \end{tabular}
 \begin{tabular}{ccc}
    \centering
    \begin{minipage}[b]{0.3\linewidth}
    \centering
    \includegraphics[keepaspectratio, width=\linewidth]{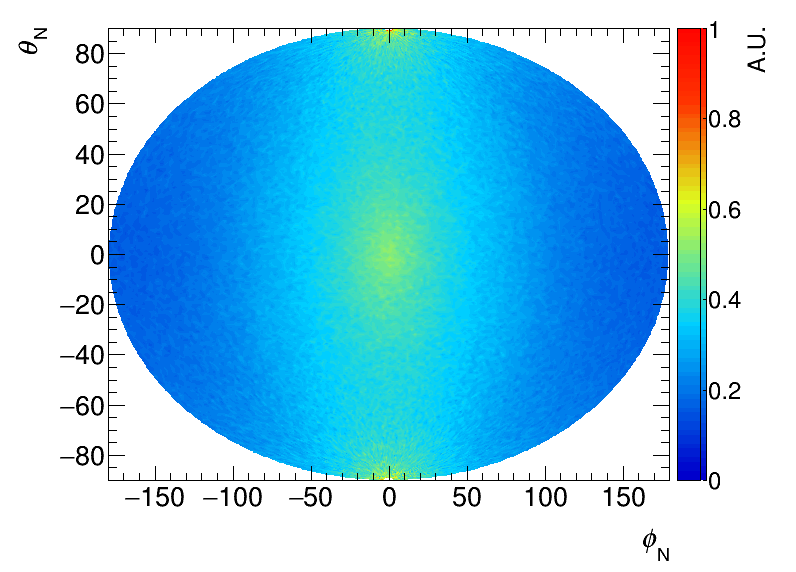}
    \caption*{NFW, $m_\chi=1$ MeV}
  \end{minipage}
  \begin{minipage}[b]{0.3\linewidth}
    \centering
    \includegraphics[keepaspectratio, width=\linewidth]{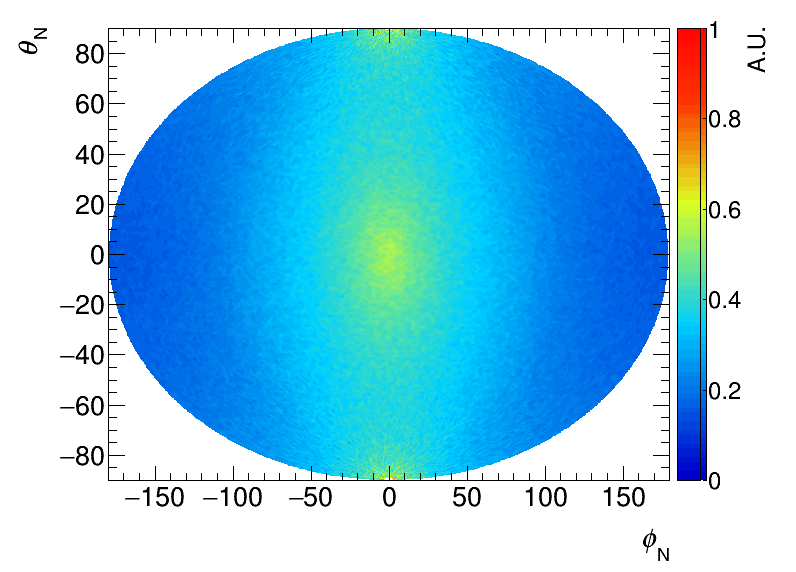}
    \caption*{Einasto, $m_\chi=1$ MeV}
    \end{minipage}
    \begin{minipage}[b]{0.3\linewidth}
    \centering
    \includegraphics[keepaspectratio, width=\linewidth]{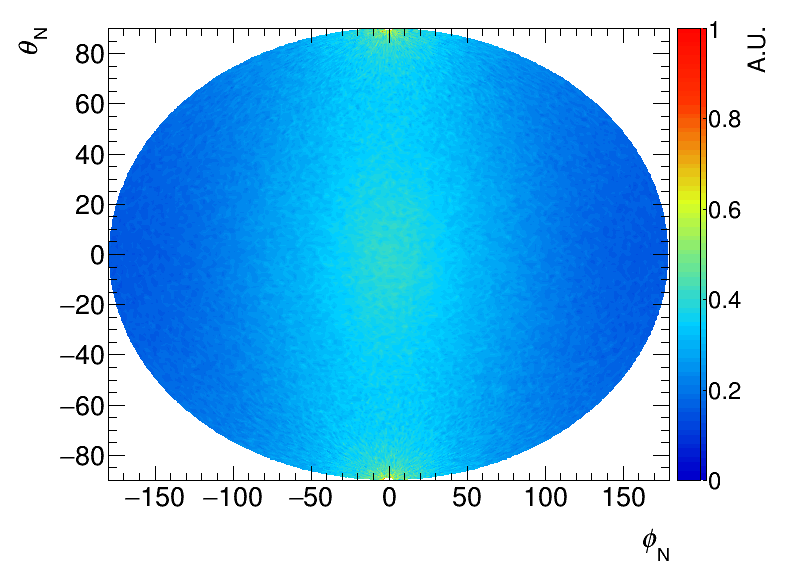}
    \caption*{PIT, $m_\chi=1$ MeV}
    \end{minipage}
 \end{tabular}
 \caption{Legend is same as Figure~\ref{Fig:skymap_p} except that the target atom in the direct detection is Ag.}
 \label{Fig:skymap_Ag}
\end{figure*}
%sky map plots end

%%%%%%%%%%%%%%%%%%%%
%Energy dependence plots
%%%%%%%%%%%%%%%%%%%%
\begin{figure*}[t]
 \begin{tabular}{ccc}
    \centering
\begin{minipage}[b]{0.3\linewidth}
    \centering
    \includegraphics[keepaspectratio, width=\linewidth]{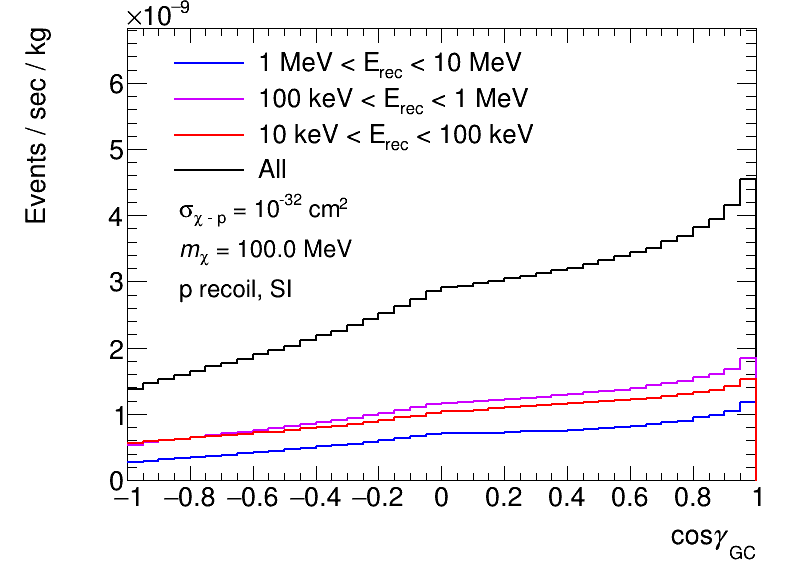}
    \caption*{NFW, $m_\chi=100$ MeV}
  \end{minipage}
  \begin{minipage}[b]{0.3\linewidth}
    \centering
    \includegraphics[keepaspectratio, width=\linewidth]{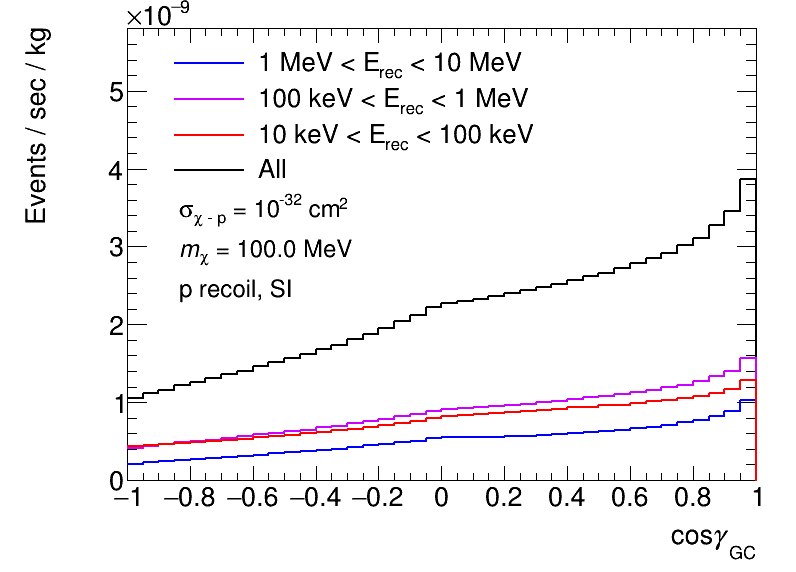}
    \caption*{Einasto, $m_\chi=100$ MeV}
    \end{minipage}
    \begin{minipage}[b]{0.3\linewidth}
    \centering
    \includegraphics[keepaspectratio, width=\linewidth]{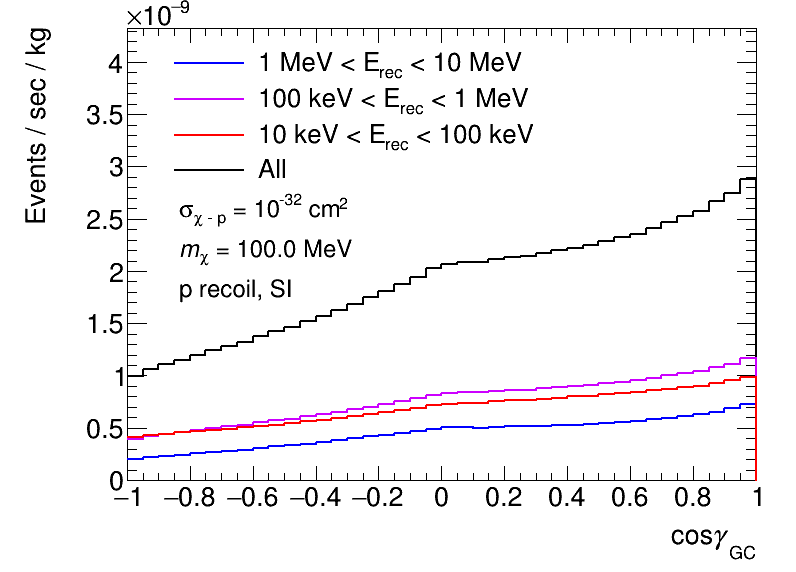}
    \caption*{PIT, $m_\chi=100$ MeV}
    \end{minipage}
 \end{tabular}
 \begin{tabular}{ccc}
    \centering
    \begin{minipage}[b]{0.3\linewidth}
    \centering
    \includegraphics[keepaspectratio, width=\linewidth]{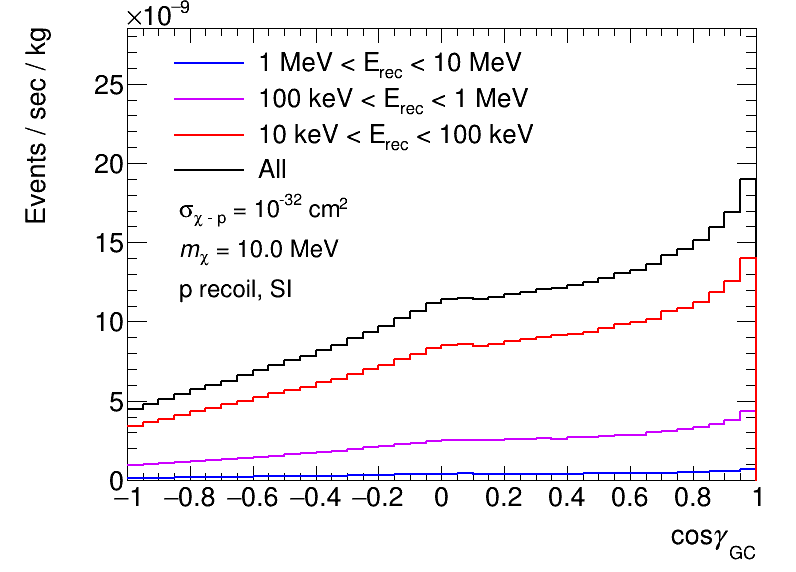}
    \caption*{NFW, $m_\chi=10$ MeV}
  \end{minipage}
  \begin{minipage}[b]{0.3\linewidth}
    \centering
    \includegraphics[keepaspectratio, width=\linewidth]{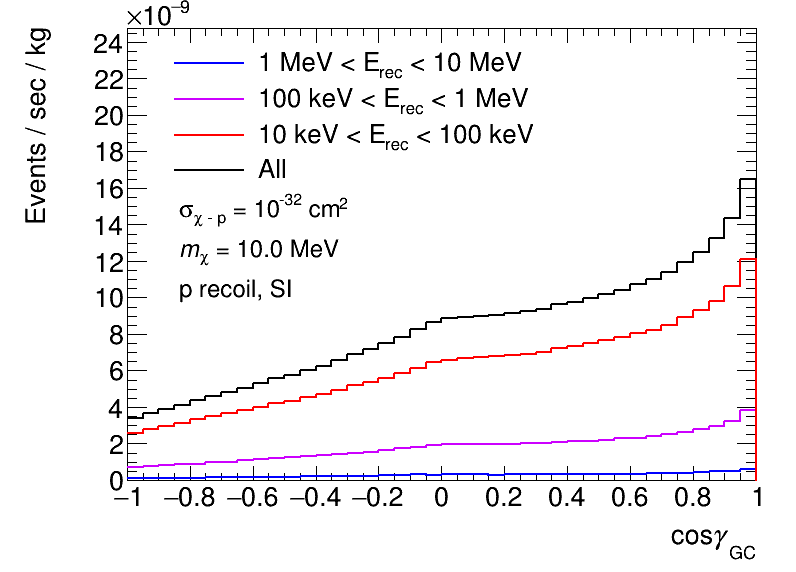}
    \caption*{Einasto, $m_\chi=10$ MeV}
  \end{minipage}
  \begin{minipage}[b]{0.3\linewidth}
    \centering
    \includegraphics[keepaspectratio, width=\linewidth]{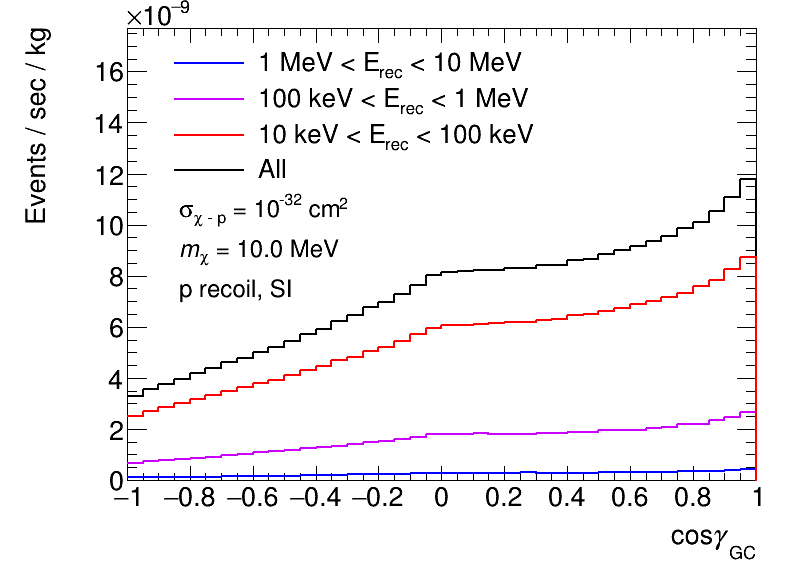}
    \caption*{PIT, $m_\chi=10$ MeV}
    \end{minipage}
 \end{tabular}
 \begin{tabular}{ccc}
    \centering
    \begin{minipage}[b]{0.3\linewidth}
    \centering
    \includegraphics[keepaspectratio, width=\linewidth]{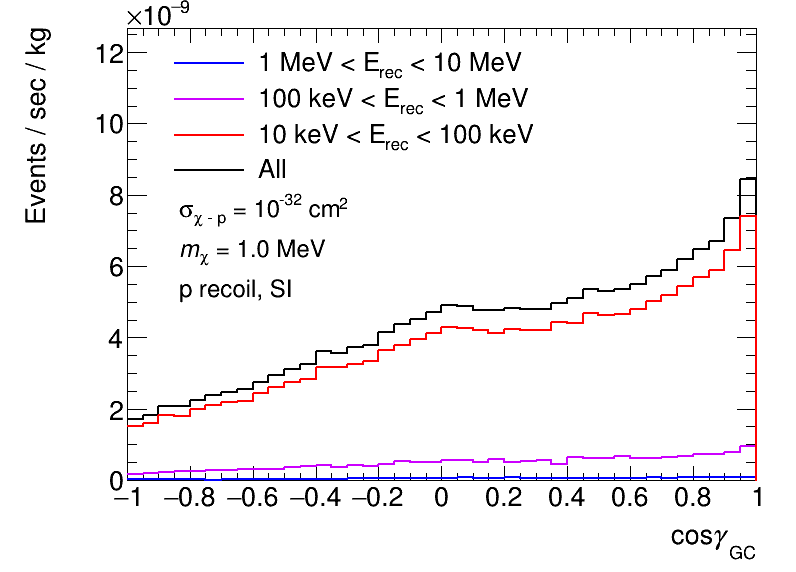}
    \caption*{NFW, $m_\chi=1$ MeV}
  \end{minipage}
  \begin{minipage}[b]{0.3\linewidth}
    \centering
    \includegraphics[keepaspectratio, width=\linewidth]{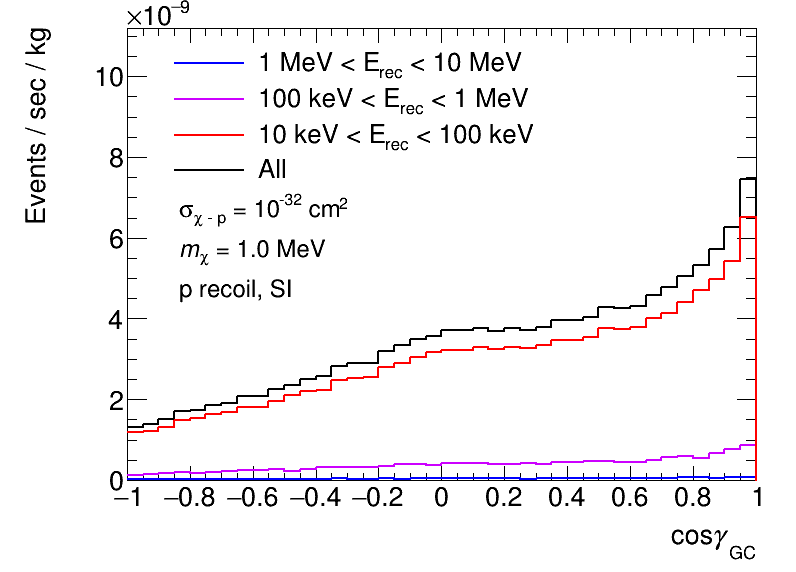}
    \caption*{Einasto, $m_\chi=1$ MeV}
  \end{minipage}
  \begin{minipage}[b]{0.3\linewidth}
    \centering
    \includegraphics[keepaspectratio, width=\linewidth]{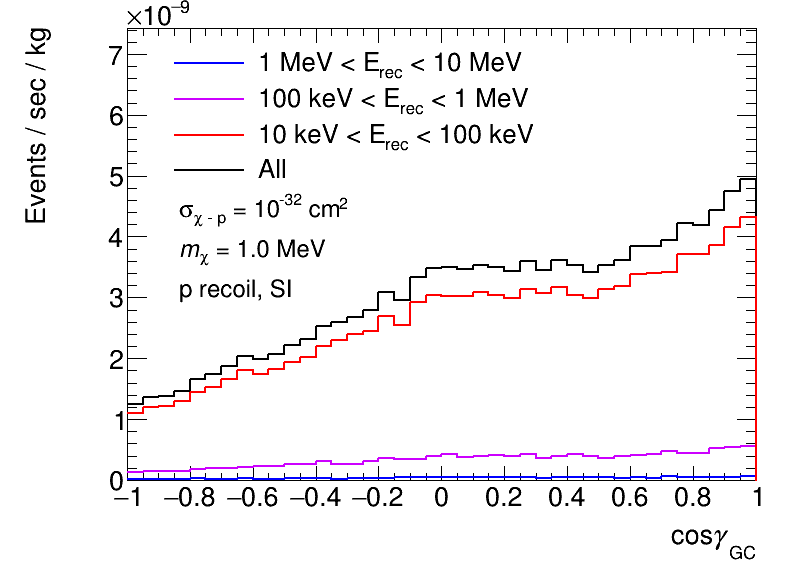}
    \caption*{PIT, $m_\chi=1$ MeV}
    \end{minipage}
 \end{tabular}
 \caption{Angular distribution of the nuclear recoil in each energy band, in the Solar system. The left column, the center column, and the right column correspond to the NFW, the Einasto, and the PIT profiles, respectively. The upper row corresponds to a CR-DM mass of 100~MeV, the middle row to 10~MeV, and the lower row to 1~MeV. The nuclear target is $p$. We assume SI interactions between the DM particle and the nuclear target.\\}
 \label{Fig:recoil_cosGCSI_p}
\end{figure*}
%%%%%%%%%%%%%%%%%%%%%%%%%%%%%%%%%%%%%%%%%%%%%%%%%%%%%%%%%%%%%%%%%%%%%%%%%%%%%%%%%
\begin{figure*}[t]
 \begin{tabular}{ccc}
    \centering
\begin{minipage}[b]{0.3\linewidth}
    \centering
    \includegraphics[keepaspectratio, width=\linewidth]{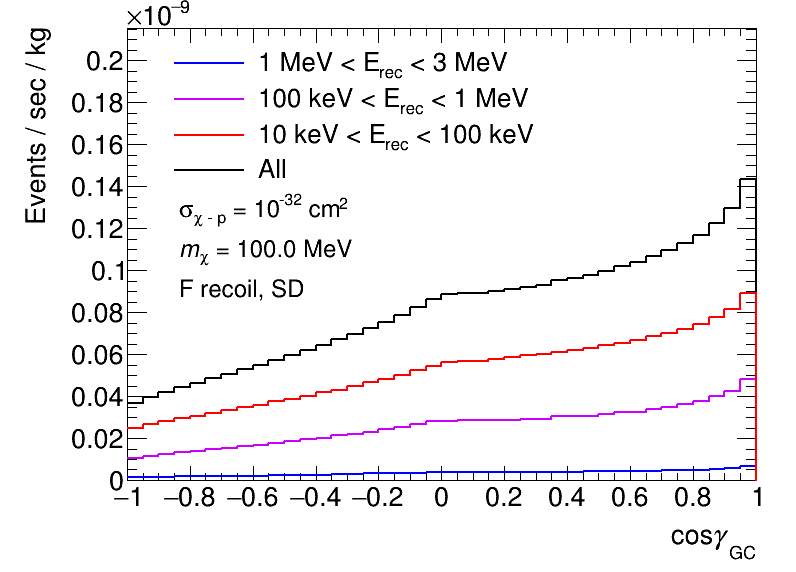}
    \caption*{NFW, $m_\chi=100$ MeV}
  \end{minipage}
  \begin{minipage}[b]{0.3\linewidth}
    \centering
    \includegraphics[keepaspectratio, width=\linewidth]{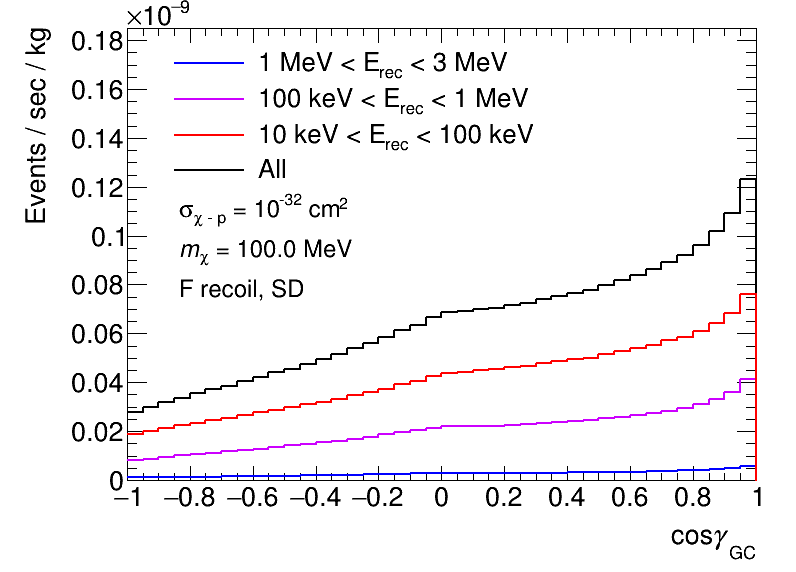}
    \caption*{Einasto, $m_\chi=100$ MeV}
    \end{minipage}
    \begin{minipage}[b]{0.3\linewidth}
    \centering
    \includegraphics[keepaspectratio, width=\linewidth]{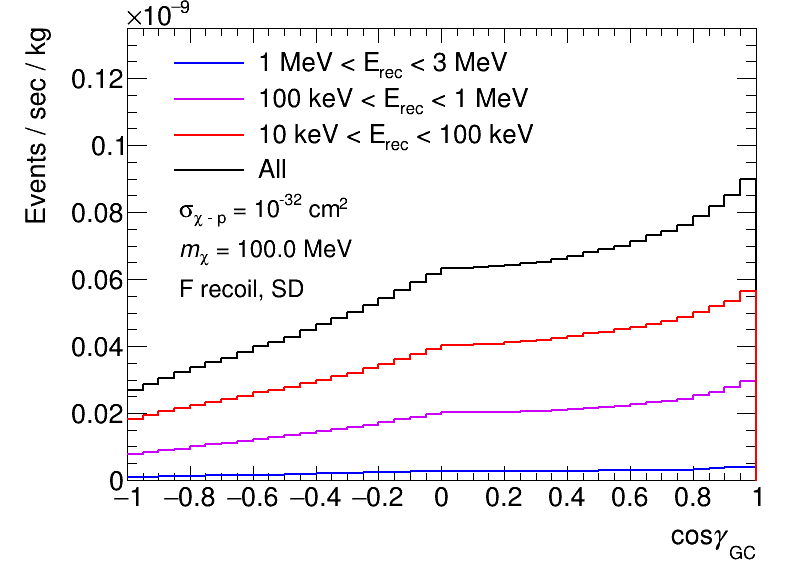}
    \caption*{PIT, $m_\chi=100$ MeV}
    \end{minipage}
 \end{tabular}
 \begin{tabular}{ccc}
    \centering
    \begin{minipage}[b]{0.3\linewidth}
    \centering
    \includegraphics[keepaspectratio, width=\linewidth]{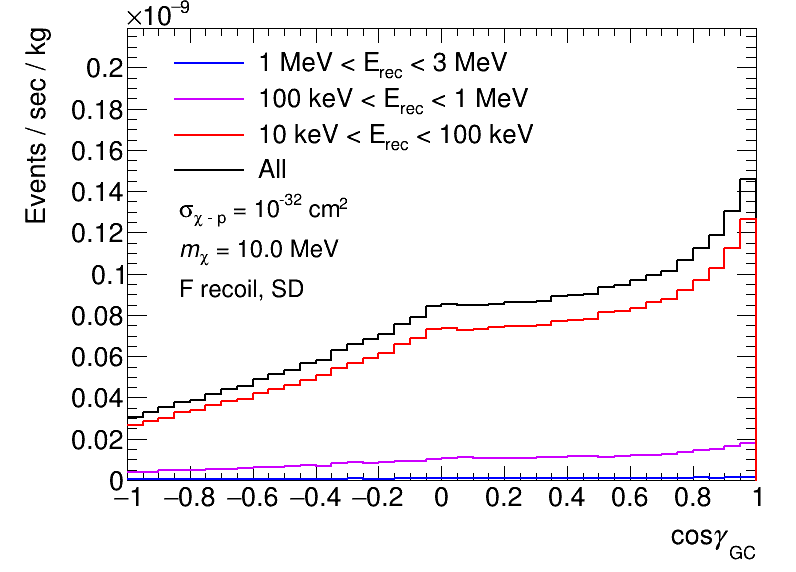}
    \caption*{NFW, $m_\chi=10$ MeV}
  \end{minipage}
  \begin{minipage}[b]{0.3\linewidth}
    \centering
    \includegraphics[keepaspectratio, width=\linewidth]{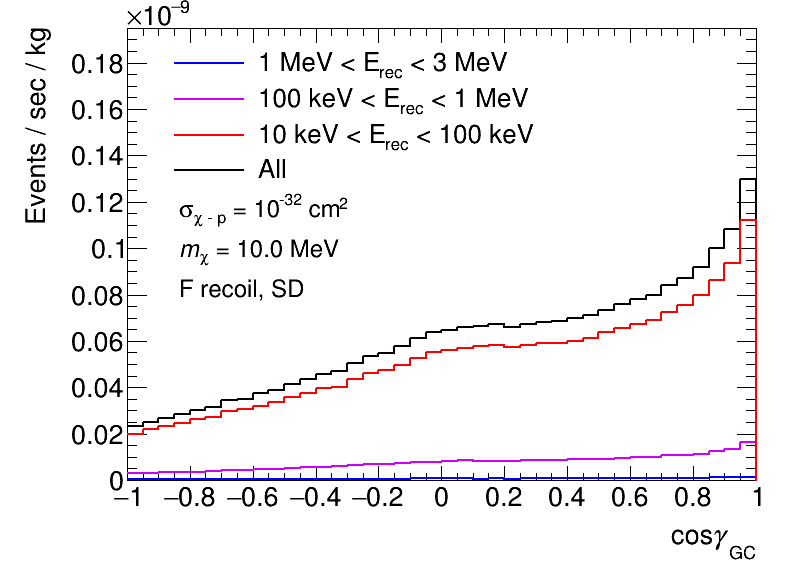}
    \caption*{Einasto, $m_\chi=10$ MeV}
  \end{minipage}
  \begin{minipage}[b]{0.3\linewidth}
    \centering
    \includegraphics[keepaspectratio, width=\linewidth]{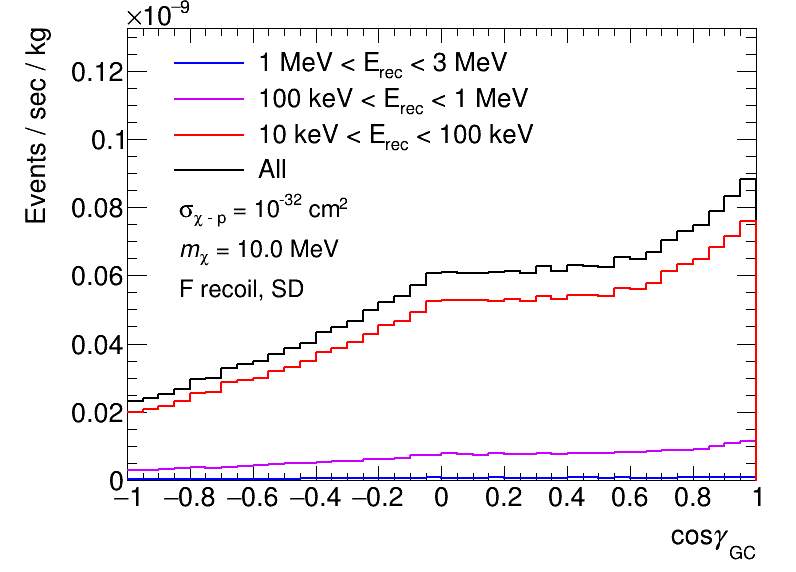}
    \caption*{PIT, $m_\chi=10$ MeV}
    \end{minipage}
 \end{tabular}
 \begin{tabular}{ccc}
    \centering
    \begin{minipage}[b]{0.3\linewidth}
    \centering
    \includegraphics[keepaspectratio, width=\linewidth]{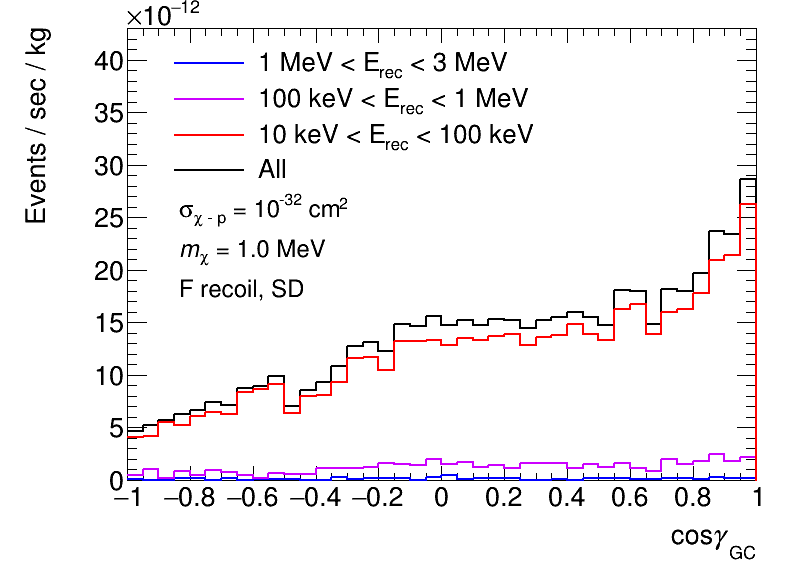}
    \caption*{NFW, $m_\chi=1$ MeV}
  \end{minipage}
  \begin{minipage}[b]{0.3\linewidth}
    \centering
    \includegraphics[keepaspectratio, width=\linewidth]{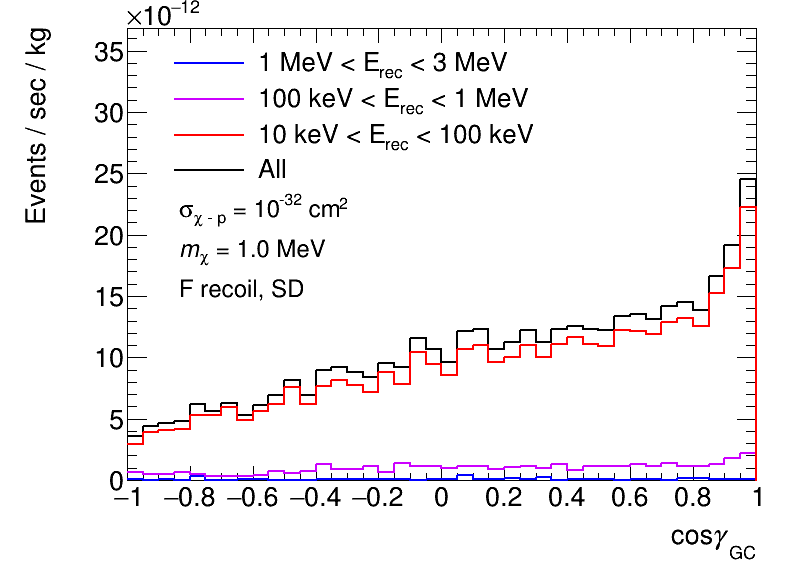}
    \caption*{Einasto, $m_\chi=1$ MeV}
  \end{minipage}
  \begin{minipage}[b]{0.3\linewidth}
    \centering
    \includegraphics[keepaspectratio, width=\linewidth]{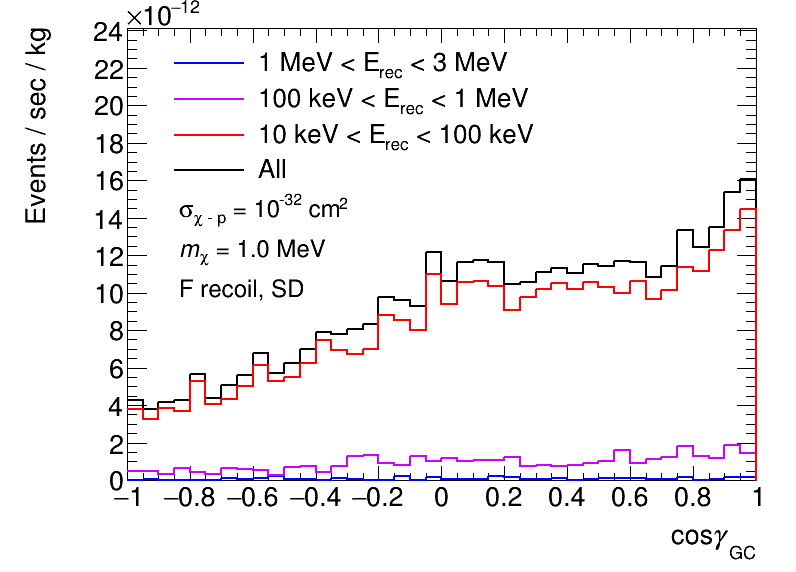}
    \caption*{PIT, $m_\chi=1$ MeV}
    \end{minipage}
 \end{tabular}
 \caption{Legend is same as Figure~\ref{Fig:recoil_cosGCSI_p} except that the target atom in the direct detection is F.}
 \label{Fig:recoil_cosGCSD_F}
\end{figure*}
%%%%%%%%%%%%%%%%%%%%%%%%%%%%%%%%%%%%%%%%%%%%%%%%%%%%%%%%%%%%%%%%%%%%%%%%%%%%%%%%%
\begin{figure*}[t]
 \begin{tabular}{ccc}
    \centering
\begin{minipage}[b]{0.3\linewidth}
    \centering
    \includegraphics[keepaspectratio, width=\linewidth]{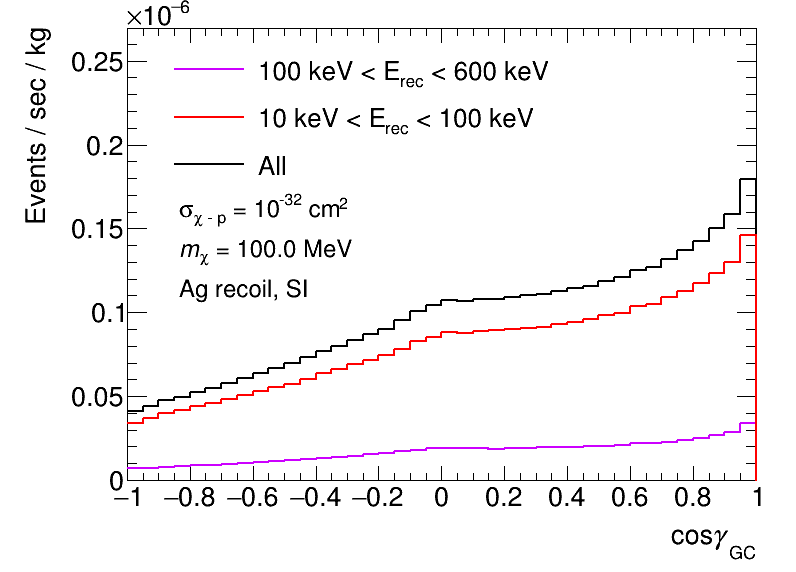}
    \caption*{NFW, $m_\chi=100$ MeV}
  \end{minipage}
  \begin{minipage}[b]{0.3\linewidth}
    \centering
    \includegraphics[keepaspectratio, width=\linewidth]{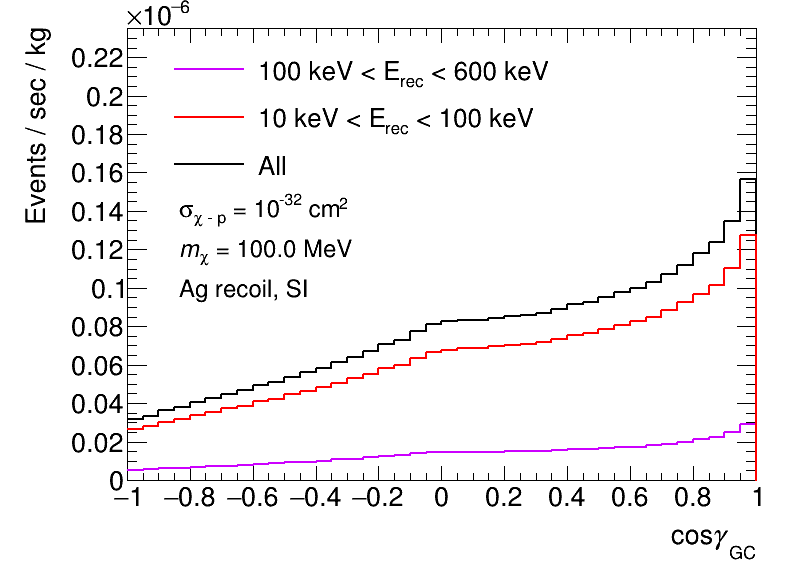}
    \caption*{Einasto, $m_\chi=100$ MeV}
    \end{minipage}
    \begin{minipage}[b]{0.3\linewidth}
    \centering
    \includegraphics[keepaspectratio, width=\linewidth]{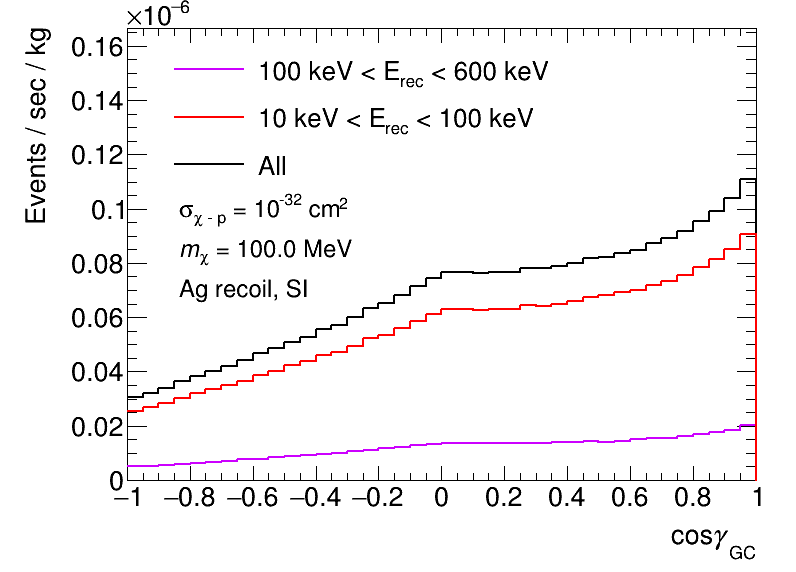}
    \caption*{PIT, $m_\chi=100$ MeV}
    \end{minipage}
 \end{tabular}
 \begin{tabular}{ccc}
    \centering
    \begin{minipage}[b]{0.3\linewidth}
    \centering
    \includegraphics[keepaspectratio, width=\linewidth]{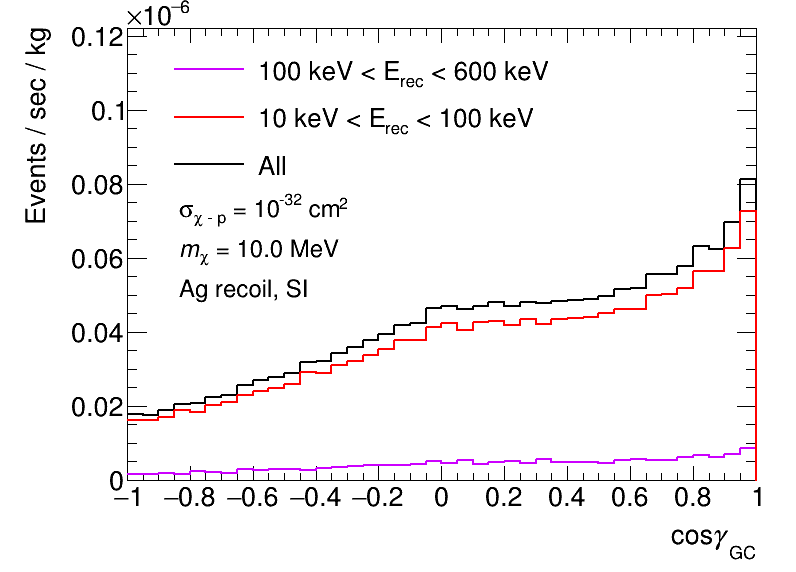}
    \caption*{NFW, $m_\chi=10$ MeV}
  \end{minipage}
  \begin{minipage}[b]{0.3\linewidth}
    \centering
    \includegraphics[keepaspectratio, width=\linewidth]{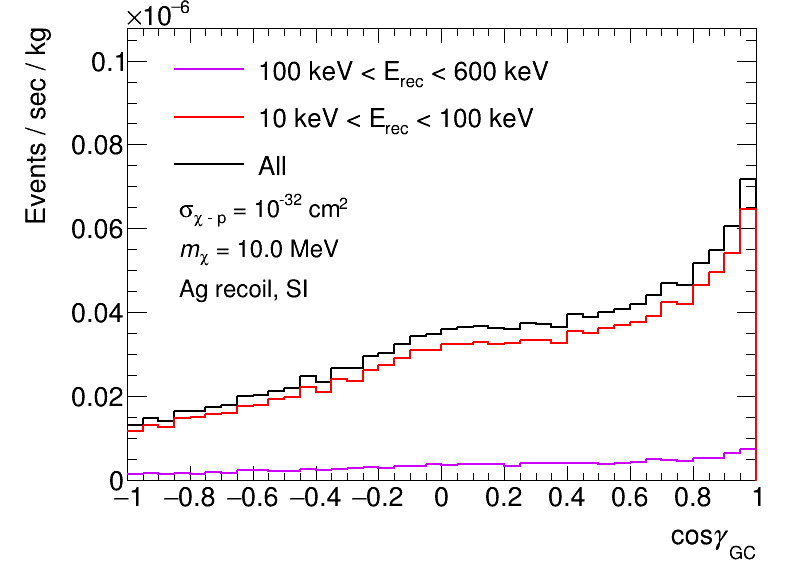}
    \caption*{Einasto, $m_\chi=10$ MeV}
  \end{minipage}
  \begin{minipage}[b]{0.3\linewidth}
    \centering
    \includegraphics[keepaspectratio, width=\linewidth]{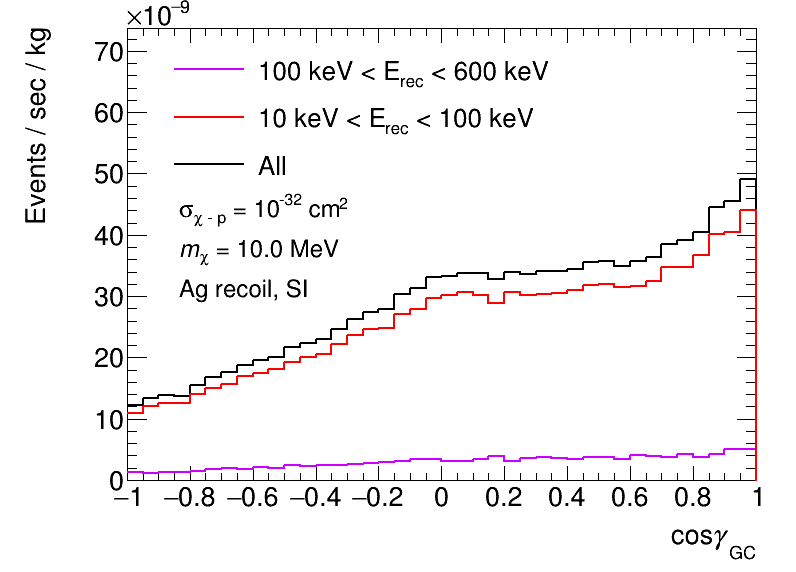}
    \caption*{PIT, $m_\chi=10$ MeV}
    \end{minipage}
 \end{tabular}
 \begin{tabular}{ccc}
    \centering
    \begin{minipage}[b]{0.3\linewidth}
    \centering
    \includegraphics[keepaspectratio, width=\linewidth]{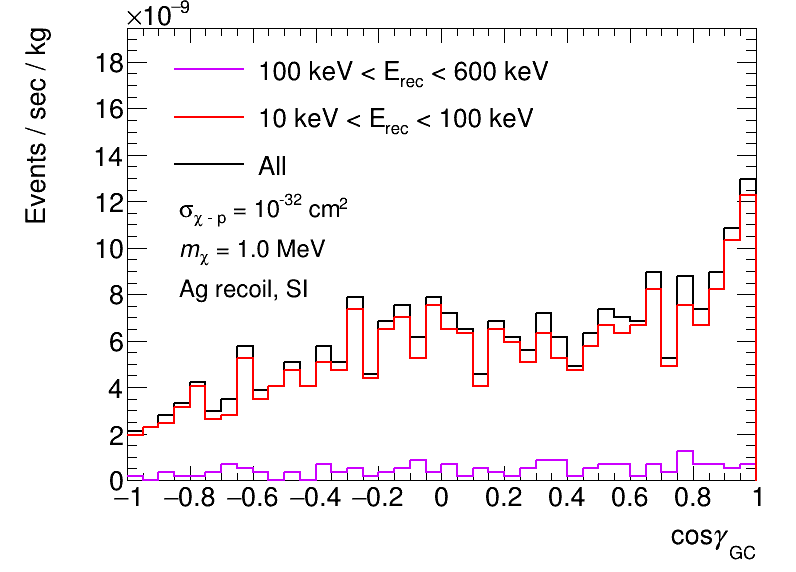}
    \caption*{NFW, $m_\chi=1$ MeV}
  \end{minipage}
  \begin{minipage}[b]{0.3\linewidth}
    \centering
    \includegraphics[keepaspectratio, width=\linewidth]{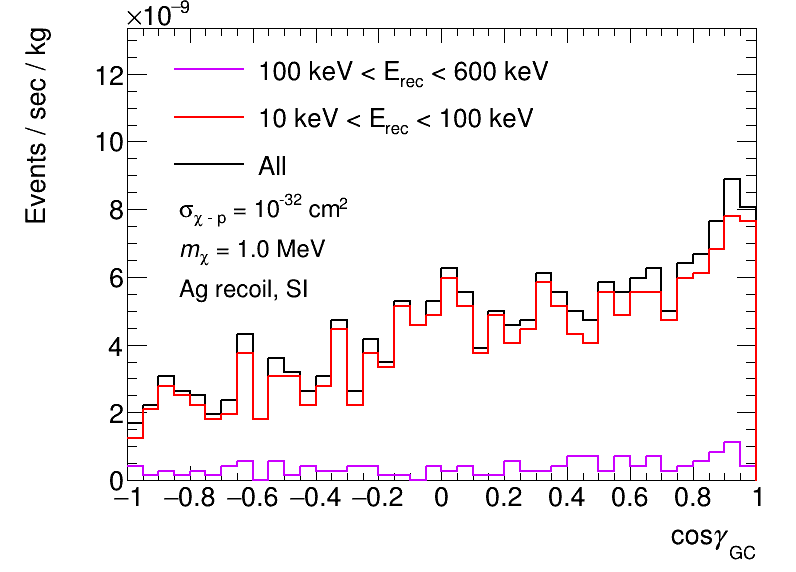}
    \caption*{Einasto, $m_\chi=1$ MeV}
  \end{minipage}
  \begin{minipage}[b]{0.3\linewidth}
    \centering
    \includegraphics[keepaspectratio, width=\linewidth]{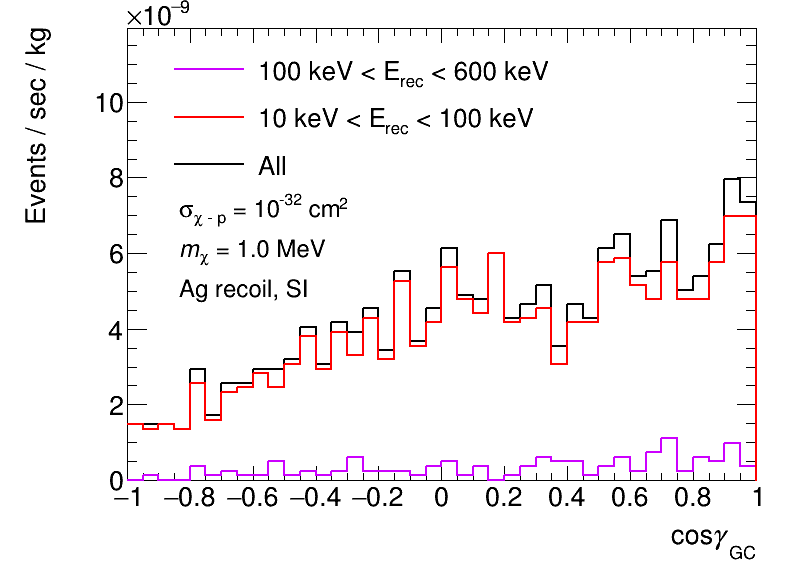}
    \caption*{PIT, $m_\chi=1$ MeV}
    \end{minipage}
 \end{tabular}
 \caption{Legend is same as Figure~\ref{Fig:recoil_cosGCSI_p} except that the target atom in the direct detection is Ag.  As mentioned in subsection \ref{subsec:energydependence}, we assume $E_\mathrm{thr}=100$ keV in the analysis of asymmetry, but data for $10 < E_R < 100$ keV are also shown here for reference.}
 \label{Fig:recoil_cosGCSI_Ag}
\end{figure*}
%energy dependence plots end

%%%%%%%%%%%%%%%%%%%%
\begin{acknowledgments}
It is a pleasure to thank S.~Matsumoto, K.~Hata, K.~Hayashi, C.~Cappiello and H.~Oshima 
%Kazumi~Hata, Shigeki Matsumoto, Kohei Hayashi, Christopher Cappiello and Hitoshi Oshima
for helpful discussions. KIN is also grateful to S.~Kime,  T.~Tomiya and N.~Ikeda for discussion on the early stages of this work. This work was supported by JSPS KAKENHI Grant Number 19H05806, 26104005, (A) 16H02189, (A) 18H03699, (C) 21K03562, (C) 21K03583, 21K13943, 22H04574 and Wesco Scientific Promotion Foundation. 
\end{acknowledgments}
%%%%%%%%%%%%%%%%%%%%

\bibliography{main}

\appendix
\section{The SI and SD cross sections}
\label{append:SIandSDcrossections}
Fluorine is the target assumed in the directional direct detection through SD interactions. 
The sensitivity to SI and SD cross sections varies by a factor $\eta_A=\sigma_{\chi-p}^\mathrm{SI}/\sigma_{\chi-p}^\mathrm{SD}$.
By evaluating the factor, the sensitivity to SD interactions in Section \ref{subsec:energydependence} can be scaled to that to SI interactions.
The cross sections of DM and nucleus scattering for the SI and SD interactions $\sigma_{\chi-N}^\mathrm{SI, SD}$ are 
\begin{align}
\sigma_{\chi-N}^\mathrm{SI}&=\sigma^\mathrm{SI}_{\chi-p}\frac{\mu_{\chi-N}^2}{\mu_{\chi-p}^2}A^2 \mathrm{[cm^2]},\\
\sigma_{\chi-N}^\mathrm{SD}&=\sigma^\mathrm{SD}_{\chi-p}\frac{\mu_{\chi-N}^2}{\mu_{\chi-p}^2}\frac{\lambda^2 J(J+1)}{0.75} \mathrm{[cm^2]},
\end{align}
respectively, where $\sigma^\mathrm{SD, SI}_{\chi-p}$ is the DM and proton scattering cross section, $\mu_{\chi-N(p)}$ is the reduced mass of CR-DM and nucleus (proton), $A$ is mass number, and $\lambda J(J+1) = 0.647$ for F. Therefore, the ratio of event number required to reach the same cross section $\sigma_{\chi-p}$, which is also denoted as $\eta_A=n_\mathrm{SI}/n_\mathrm{SD}$, is estimated as
\begin{align}
\eta_A&=\frac{0.75 A^2}{\lambda^2 J(J+1) }\nonumber\\
&\simeq 418.
\end{align}
\end{document}